\documentclass[journal]{IEEEtran}

\ifCLASSINFOpdf
\else
\fi

\usepackage{xr}

\usepackage{amssymb}
\usepackage{amsmath}
\usepackage{cite}
\usepackage{graphicx}
\usepackage{subfig}
\usepackage{epstopdf}
\usepackage{algorithm}
\usepackage{algorithmic}
\usepackage{tikz}
\usepackage{lscape}
\usepackage{graphicx}
\usepackage{fancyhdr}
\usepackage{tikz}
\usepackage{epstopdf}
\usepackage{array}
\usepackage{multirow}
\usepackage{amsmath,amsthm}
\usepackage{footnote}
\usepackage{supertabular}
\usepackage{mathtools}
\usepackage{amssymb}
\usepackage{longtable}
\usepackage{nomencl}
\usepackage[nonumberlist,acronym]{glossaries}
\usepackage{algorithmic}
\usepackage{color}
\usepackage{algorithm}
\usepackage{enumerate}
\usepackage{datatool}
\usepackage{amsmath}
\usepackage{nomencl}%
\usepackage{amsfonts}
\usepackage{authblk}
\providecommand{\U}[1]{\protect\rule{.1in}{.1in}}
\usetikzlibrary{arrows,automata}
\usetikzlibrary{automata}

\ifCLASSINFOpdf
\else
\fi
\ifCLASSOPTIONcompsoc
\else
\fi
\begin{document}

\title{Correlated Eigenvalues Optical Communications}

\author[1]{Wen Qi Zhang$^{*}$\thanks{$^{*}$These authors have contributed similarly to the paper}}
\author[2]{Tao Gui$^{*}$} 
\author[3]{Qun Zhang}
%\author[2]{Xuebing Zhang}
\author[4]{Chao Lu}
\author[1,5]{Tanya M. Monro}
\author[3]{Terence H. Chan}
\author[2]{Alan Pak Tao Lau}
\author[1,5]{Shahraam Afshar V.}
\affil[1]{Laser Physics and Photonic Devices Laboratories, School of Engineering, University of South Australia,  Australia}
\affil[2]{Department of Electrical Engineering, Photonics Research Center, The Hong Kong Polytechnic University} 
\affil[3]{Institute for Telecommunications Research, University of South Australia, Australia}
\affil[4]{Department of Electronic and Information Engineering, Photonics Research Center, The Hong Kong Polytechnic University}
\affil[5]{Institute for Photonics and Advanced Sensing, School of Physical Sciences, The University of Adelaide, Australia}

% \authorcr Email: {\tt \{uid1, uid2\}@usc.edu}
% \author{Wen Qi Zhang$^{1}$, Qun Zhang$^{2}$, Kashif Amir$^{1}$, Tao Gui$^{3}$, Xuebing Zhang$^{4}$,\\
% Alan Pak Tao Lau$^{3}$, Chao Lu$^{4}$, Terence H. Chan$^{2}$, and Shahraam Afshar V.$^{1,5}$}

% \IEEEauthorblockA{Department of Whatever,
% Whichever University\\
% Wherever\\}
% \address{$^{1}$ Laser Physics and Photonic Devices Laboratories, School of Engineering, University of South Australia,  Australia;\\}
% %
% $^{2}$Institute for Telecommunications Research, University of South Australia, Australia;\\
% %
% $^{3}$
% Institute for Photonics and Advanced Sensing, School of Physical Sciences, The University of Adelaide, Australia.
% 
%\email{e-mail address}
\maketitle

\begin{abstract}
There is a fundamental limit on the capacity of fibre optical communication system (Shannon Limit). This limit can be potentially overcome via using Nonlinear Frequency Division Multiplexing. Dealing with noises in these systems is one of the most critical parts in implementing a practical system. 

In this paper, we discover and characterize the correlations among the NFT channels. It is demonstrated that the correlation is universal (i.e., independent of types of system noises) and can be exploited to maximize transmission throughput. We propose and experimentally confirm a noise model showing that end-to-end noise can be modeled as the accumulation of noise associated with each segment of optical communication which can be dealt with independently.  Also, each point noise can be further decomposed into different components, some of which are more significant (and even dominating) than others. Hence, one can further approximate and simplify the noise model by focusing on the significant component.

\end{abstract}

%\markboth{Journal of \LaTeX\ Class Files,~Vol.~14, No.~8, August~2015}{Shell
%\MakeLowercase{\textit{et al.}}: Bare Demo of IEEEtran.cls for IEEE Journals}

%\begin{IEEEkeywords}
%IEEE, IEEEtran, journal, \LaTeX, paper, template.
%\end{IEEEkeywords}

\newtheorem{df}{Definition} \newtheorem{thm}{Theorem}
\newtheorem{prop}{Proposition} \newtheorem{lemma}{Lemma}
\newtheorem{example}{Example} \newtheorem{cor}{Corollary}
\newtheorem{rem}{Remark} \newtheorem{conjecture}{Conjecture}

\newcommand{\tc}[1]{{\color{red} #1}}

\IEEEpeerreviewmaketitle

\def\imag{{j}}

\section{Introduction}

Data traffic has been growing at a rate of more than 60\% per year \cite{2010T}.
Such astronomical growth has sparked an urgent need to significantly increase
the network transmission capacity, posing a critical technical
challenge for system designers. 
One main fundamental challenge to further
enhance data transmission bandwidth is to manage fibre nonlinearities\cite{b4}. 

Signal propagation across an optical fibre is governed by the nonlinear Schr\"{o}dinger equation. The channel is nonlinear, unlike other typical transmission media such as copper wires and radio waves \cite{b4}. 
Traditionally, fibre nonlinearities are often regarded as channel
impairments, and hence should be eliminated or mitigated.  Instead of dealing with fibre nonlinearities directly,  existing schemes are often based on a ``flawed'' approach in that they apply ``off-the-shelf'' methods originally developed for classical linear time-invariant radio frequency channels (typically with additive white Gaussian noise). This approach ignores the detail of the underlying fibre physics, and attempts to draw loose analogies between macroscopic channel impairments (e.g. dispersion caused by a linear multipath channel) encountered in microwave channels with those in optical channels (e.g. dispersion due to wavelength-dependent refractive index, fibre geometry or nonlinearities). In essence,  nonlinearities are assumed to be  weak and hence can be treated and suppressed as  small perturbations~\cite{2010JLT/EKWFG,2008JLT/IK}.

The underlying premise behind this perspective
is that signals are processed often in the time domain and/or the (linear)
frequency domain (where signals are obtained by applying linear transformation
such as Fourier transform on the time-domain signals). However, fibre
nonlinearities cannot be completely eliminated by invoking these linear signal
processing techniques, leading to undesirable inter-symbol interference (ISI) and
inter-channel interference (ICI) \cite{2008JLT/IK}. As a result, it was noted that fibre nonlinearities can impose a fundamental limit (known as linear Shannon limit) on the data transmission capacity \cite{2010JLT/EKWFG}.

% \tc{ Delete this? 
% Specifically, as fibre nonlinearity, inter-channel interference is inevitable, which reduces the performance of the channel, when trying to do the multiplexing to allow multiuser communication, such as orthogonal frequency division multiplexing (OFDM) and wavelength division multiplexing (WDM), etc. 
% }

A different paradigm to the problem has received a lot of attentions in the past few years. In this paradigm, fibre nonlinearity and dispersion effect are merely seen as ordinary physical characteristics needed to be managed directly, rather than simply evading them as disadvantages~\cite{1993HN,frank1,DBLP:journals/corr/abs-1204-0830,Frankpt3}. In particular, their approaches are based on the use of nonlinear Fourier transform (NFT), or direct scattering transform \cite{1972JETP/ZS,b3,b1}.
Higher order dispersion effects are often ignored, and the linear loss term is assumed to be perfectly compensated by the distributed Raman amplification (DRA). 

%to investigate an integrable system charaterising the propagation of the normalized complex electrical field in an optical fibre using the 

Mathematically, the NFT provides a systematic method for solving the class of
integrable nonlinear Schr\"{o}dinger equation, whereas in engineering
perspective, NFT can ``decompose'' the nonlinear fibre channel into multiple independent subchannels in the (nonlinear) spectral domain. To a great extent, it  mirrors the widely used wavelength-division multiplexing (WDM), a technology which multiplexes multiple optical carrier signals and transmitted in a single optical fibre. The fundamental difference between WDM and NFT based approach is in how the ``\emph{modes}'' or ``\emph{subchannels}''  are defined. 

Roughly speaking, WDM employs Fourier Transform (FT) such that each wavelength (or its corresponding linear frequency) is essentially a ``transmission mode''. When signal-to-noise ratio (SNR) is low and nonlinearities are not severe,  interference among these modes are negligible. However, as data rate (and also signal power) increases, nonlinearities become significant and the transmission modes defined by FT can now significantly interfere with each other. This significantly limits the performance of the fibre-optic communications systems, especially in long-haul transmissions. In \cite{frank1}, NFT was used instead of FT, such that the resulting nonlinear normal modes will not interfere with each other even in the presence of nonlinear effects.
This idea of decoupling a nonlinear channel into multiple independent subchannels plays the central role in NFT based communications. As a result of the channel decomposition, one can  separately  design communications for each individual subchannel, and hence greatly reduce the system complexity. Also, inter-channel interference is eliminated with a proper
allocation of the (nonlinear) spectrum to users at least in the noise free
scenario. This scheme  is  called nonlinear frequency
division multiplexing (NFDM) \cite{frank1}.

The development of NFT-based transmission systems is only in its infancy stage at the moment. Some preliminary experimental works have already been done to demonstrate the concepts.
The spectrum of a time domain signal, after applying the NFT, is composed by
discrete and continuous spectrum.  Both continuous \cite{frank1,Prilepsky2014,Le201426720,Le2016,Prilepsky201324344} and discrete \cite{Dong2015,Terauchi2013, Maruta2015,Meron20124458,Oda2004587,Bulow20151433} spectra have been proposed for optical transmission systems. Using a discrete 1-, 2-, 3-eigenvalue configuration together with on-off keying, Dong {\em et. al}, \cite{Dong2015} have achieved 1.5 Gbps transmission over 1800 km. In \cite{Aref2015} a two-eigenvalue signal together with QPSK modulation has been used to achieve 4 Gbps transmission rate over 640km. Optical transmission using continuous spectrum, has recently been demonstrated by Le {\em et. al} \cite{Le2016}, where a 120 bits/burst trasmission over 7344 km has been achieved.    

As mentioned above, NFT based methods lead to channel decomposition with zero inter-channel interference.   
Unfortunately, perfect channel decomposition is only theoretically possible in the absence of noise. In practice, noise can be generated in the transmitter and the receiver (e.g., quantization, clipping), and also during propagation in the fiber (e.g., due to inline or point amplification). These noises will induce correlations among individual subchannels, affecting the capacity of the transmission system. Despite the crucial role of noise in determining the actual capacity of an NFT-based transmission system, effects of noise on NFT continuous and discrete spectra have only been studied in limited cases. Zhang {\em et. al.} \cite{Zhang2015}, have studied the effect of propagation noise (with Gaussian distribution) on the spectral amplitudes and the discrete eigenvalue of the channel output when the input is a fundamental soliton. In addition, Derevyanko {\em et. al.}, \cite{Derevyanko2016} have developed an approximated noise (with Gaussian distribution) model for continuum spectra NFT based transmission and estimated a lower bound for the capacity. Based on their model, they find the noise properties of NFT continuous spectra after propagating through a fiber in presence of noise.  

In this paper,  we investigate the noise properties of optical communications systems based on input optical pulses with discrete NFT eigenvalues. There are four main aspects in our contributions. \emph{First}, 
we demonstrate that the discrete eigenvalues of a signal propagating through the network are correlated, regardless of the different types of noise that have been introduced at different stages of signal preparation, propagation, and detection. 
% The correlation of discrete eigenvalues is demonstrated experimentally in a fiber optic communication setup including a $400$ km standard single mode fiber and is also confirmed by simulating pulse propagation in similar optical fiber. 
We show that such correlation properties can be used to maximize the transmission throughput so that input signal constellation can be optimized to support high data transmission rate. 

\emph{Second},
% motivated by the idea that a nonlinear function can be approximated by a piece-wise linear function, 
we propose and experimentally confirm a noise model, in which deterministic (due to nonlinearity and dispersion) and stochastic (due to all noises) effects on signals can be separately evaluated in a similar fashion as linear and nonlinear processes in Split Step Method. 

\emph{Third}, it is shown that the noise effects do not accumulate and noise associated with each segment of optical communication can be dealt independently.  
% offer an analytically framework to derive a model for the correlations among discrete eigenvalues. The resulting framework is much simpler that decouples the stochastic noise (e.g., caused by inline amplificaiton) and deterministic nonlinearities and dispersion.
%
%and allows the elimination of coupling effects of noises and signals in NFT domain (Is this true Terence???). 

\emph{Finally}, as a result of second and third properties, we identify and demonstrate that the effect of noise can be decomposed into different components and some noise components are more significant than others. This suggests that one can further approximate and simplify noise effects by focusing on those significant noise components. 
 
%
%\bigskip

%%%%%%%%%%%%%%%%%%%%%%%%%%%%%%%%%%%%
\section{Basic Principle}

%\tc{change all fibre length to fancy L, remove subscript n, get ride of }

The noisy signal evolution across an optical fibre is often modelled as the
following stochastic nonlinear Schr\"{o}dinger equation (SNLSE)
\begin{multline}
\frac{\partial A(s,l)}{\partial l}-\frac{\imag \beta_{2}}{2}\frac{\partial
^{2}A(s,l)}{\partial s^{2}}+\frac{\alpha}{2}A(s,l) =\\
- \imag\gamma|A(s,l)|^{2}A(s,l)+\imag \kappa N(s,l),\\
\qquad0\leq l\leq\mathfrak{L}\ \mathrm{km},\label{ch3.1.3}%
\end{multline}
where $\imag=\sqrt{-1}$. 
The function $A(s,l)$ is the complex envelope of the signal propagating along the fibre. 
The parameter $\beta_{2}$ is the group velocity
dispersion (GVD) coefficient. The GVD coefficient for silica fibres is
$\beta_{2}=-2\times10^{-23}$ s$^{2}\cdot$km$^{-1}$ when the input wavelength
is $1.55$ $\mathrm{\mu m}$. The parameter $\alpha$ is called attenuation
coefficient which describes the (linear) loss effect, and $\gamma$ is the
nonlinear coefficient. The positive real number $\mathfrak{L}$ denotes the
length of the optical fiber. The term $\imag \kappa N(\tau,l)$ represents the
optical noise field, which could be modelled as a zero mean circularly
symmetric complex white Gaussian noise process \cite{b41,1994M} with
\begin{equation}
\label{ch3.1.2}{\mathrm{E}\left[ N(s,l)N^{*}(s^{\prime},l^{\prime})\right]
=\delta(s-s^{\prime})\delta(l-l^{\prime}),}%
\end{equation}
where we use ``$*$'' to denote the complex conjugate, and $\delta(x)$ means
Dirac delta function, and $\kappa$ 
%^{2}=\alpha h\nu_{s}K_{T}$. 
is a coefficient that determines the strength of the noise.
% The parameters
% are summarised in Table \ref{ch1.table1.1} obtained from
% \cite{DBLP:journals/corr/abs-1302-2875,2010JLT/EKWFG}. \begin{table}[t]
% \caption{Fibre Parameters}%
% \label{ch1.table1.1}%
% \renewcommand{\arraystretch}{1.3} \centering
% \begin{tabular}
% [c]{|c|c|c|}\hline
% Symbols & Values & Explanations\\\hline
% $\alpha$ & $0.046$ km$^{-1}$ & Fibre loss (0.2 dB/km)\\
% $h$ & $6.626\times10^{-34}$ J$\cdot$s & Planck's constant\\
% $\nu_{s}$ & $193.55$ THz & Centre frequency\\
% $K_{T}$ & $1.13$ & Photon occupancy factor \\
% $\gamma$ & $1.27$ W$^{-1}\cdot$km$^{-1}$ & Nonlinear parameter\\\hline
% \end{tabular}
% \end{table}
After applying the following variable transformations
\begin{equation}
\label{ch2.x1.1}{q=\frac{A}{\sqrt{P}},\qquad t=\frac{s}{T},\qquad
z=\frac{l}{  \mathfrak{L} },}%
\end{equation}
where
\begin{equation}
\label{ch2.x1.2}
P=\frac{2}{\gamma \mathfrak{L}},\qquad T=\sqrt{\frac
{|\beta_{2}|\mathfrak{L}}{2}},
\end{equation}
we obtain the normalised
SNLSE
\begin{equation}
\label{1.3}{\imag q_{z}(t,z)=q_{tt}(t,z)+2|q(t,z)|^{2}q(t,z)+\imag \epsilon G(t,z),}%
\end{equation}
where the noise $\epsilon G(t,z)$ is a zero mean circularly symmetric complex
white Gaussian noise with power spectral density $\epsilon^{2}=\frac{\gamma
}{\sqrt{2|\beta_{2}|}}\kappa^{2}$.
Under the model \eqref{1.3}, the effect of noise in soliton parameters was
studied. In particular, the statistics of the eigenvalue was reported in
\cite{b30,b31}, and the arrival time jitter, namely Gordon-Haus effect, was
studied in the celebrated paper \cite{1986GH}. The research about soliton
transmission control, regarding the issue of timing jitter, can be found in
\cite{1991MMHL,1992MGE,1991NYKS,1995M}. The Gordon-Mollenauer effect,
referring to the soliton phase jitter, was investigated in \cite{1990GM}, and
the work about its statistics in solitonic dispersion phase shift
keying systems was studied in \cite{1999HPGR,2000HPGR,2001HPGR,2002MX}.

Let $L$ be an operator on $q(t,z)$ where 
\begin{equation}
\label{ch3.2.3}{ L = \imag \left(
\begin{array}
[c]{ccc}%
\frac{\partial}{\partial t} & -q(t,z) & \\
-q^{*}(t,z) & -\frac{\partial}{\partial t} &
\end{array}
\right) , }%
\end{equation}

The eigenvalues  of the operator $L$ are invariant in $z$ as the
signal $q(t,z)$ propagates through the fibre. In the definition of the NFT, we
suppress the variable $z$ because it is only useful when we need to derive the
spatial signal propagation through an optical fibre. Throughout this paper, we
assume that $q(t)\in L^{1}(\mathbb{R})$ and $q(t)\rightarrow0,\ t\rightarrow
\infty.$

The NFT of a signal $q(t)$ is defined via the spectral analysis of the
operator $L$. Specifically, we need to solve the eigenvalue problem
\[
Lv=\lambda v
\]
at first, which is equivalent to the ordinary differential equation (ODE)
\begin{equation}
\label{ch3.2.7}{v_{t}=\left(
\begin{array}
[c]{ccc}%
- \imag\lambda & q(t) & \\
-q^{*}(t) & \imag \lambda &
\end{array}
\right) v}%
\end{equation}
called the scattering problem. Using  boundary conditions 
\begin{equation}
\label{ch3.2.9}{\lim_{t\rightarrow-\infty}\left| v(t,\lambda)-\left(
\begin{array}
[c]{c }%
1   \\
0  
\end{array}
\right) e^{- \imag\lambda t}\right| =0,}%
\end{equation}
we obtain a solution  of  \eqref{ch3.2.7}. 

Let $v(t,\lambda) = [v_{1} (t,\lambda), v_{2} (t,\lambda)]^{\top}$.
The coefficients $a(\lambda)$ and $b(\lambda)$ are called scattering data,
which can be obtained by calculating
\begin{equation}
\label{ch3.2.12}{a(\lambda)=\lim_{t\rightarrow\infty}v_{1} %
(t,\lambda)e^{ \imag  \lambda t},}%
\end{equation}
and
\begin{equation}
\label{ch3.2.13}{b(\lambda)=\lim_{t\rightarrow\infty}v_{2} %
(t,\lambda)e^{- \imag\lambda t}.}%
\end{equation}

%\tc{For discrete eigenvalues, should we use $\lambda_k$ instead of $\lambda_{k}$?}

The nonlinear Fourier transform of a function $q(t)$ is defined with the help
of the scattering data.
The NFT of a signal $q(t)$ is composed of its spectrum and the corresponding
spectral amplitudes. The spectrum is composed of the discrete and continuous
spectrum. The discrete spectrum is a set of isolated complex points called
(discrete) eigenvalues, which are zeros of the scattering data $a(\lambda)$ on
the upper half complex plane $\mathbb{C}^{+}\triangleq\{c\in\mathbb{C}%
:\ \mathrm{Im}(c)>0\}$. The continuous spectrum is the real line $\mathbb{R}$.
The corresponding spectral amplitudes are defined as follows. The discrete
spectral amplitude subject to an eigenvalue $\lambda_{k}\in\mathbb{C}^{+}$ is
\begin{equation}
\label{ch3.2.14}{Q^{(d)}(\lambda_{k})=\frac{b(\lambda_{k})}{a^{\prime}(\lambda_{k}%
)},\qquad k=1,2,\ldots,N,}%
\end{equation}
where $a^{\prime}(\lambda_{k})\triangleq\frac{{\mathrm{d}} a(\lambda
)}{{\mathrm{d}} \lambda}\Big|_{\lambda=\lambda_{k}}$, and $N$ is the number of
the zeros of $a(\lambda)$. The continuous spectral amplitude is defined as
\begin{equation}
\label{ch3.2.15}{Q^{(c)}(\lambda)=\frac{b(\lambda)}{a(\lambda)},}%
\end{equation}
where $\lambda\in\mathbb{R}$.

It is well known that the spectrum of the signal keeps invariant as a signal
propagates through an optical fibre in the noise free case. The spatial
evolution of the spectral amplitudes are summarised as follows:
\begin{equation}
\label{ch3.2.16}{Q^{(c)}(\lambda,z)=Q^{(c)}(\lambda,0)e^{-4 \imag  \lambda^{2}z},}%
\end{equation}
and
\begin{equation}
\label{ch3.2.17}{Q^{(d)}(\lambda_{k},z)=Q^{(d)}(\lambda_{k},0)e^{-4 \imag  \lambda_{k}%
^{2}z},\qquad k=1,2,\ldots,N,}%
\end{equation}
where $Q^{(c)}(\lambda,z)$ and $Q^{(d)}(\lambda_{k},z)$ are respectively a
continuous and a discrete spectral amplitude at position $z$, and $z>0$, and $-4 \imag  \lambda_{k}$ is the {\em NFT channel gain coefficient} and $e^{-4 \imag  \lambda_{k}^{2}z}$ as the channel gain.

%%%%%%%%%%%%%%%%%%%%%%%%%%%%%%%%%%%%%%
%\section{\bigskip Results}
\section{{\bf Results:} Eigenvalue correlation}
%\subsubsection{Positive correlation}
Pulse propagation experiments and simulations were carried out using pulses with only
two discrete eignevlues ranging from $\lambda_{1}=0.3 \imag$ to $0.75 \imag$ and
$\lambda_{2}=0.9 \imag$ to $1.35 \imag$ in steps of $0.15 \imag.$ 

%\tc{(Alan: can you have a look at this section and see if the descriptions on the experimental set up is good. }
Fig. 1 shows the experimental setup for the eigenvalue correlation
transmission system. The transmitter (TX) comprises a 92-GSa/s arbitrary
waveform generator (AWG) providing a drive signal for an IQ modulator which
generates 1-GBd optical soliton pulses train in a single polarization. The
outputs of the modulator are amplified and launched into a fiber recirculating
loop. Since the theory of NFT is based on the integrability property of the
lossless nonlinear Schr\"{o}dinger equation, a short span of 50km NZ-DSF fiber (with $\alpha$ = 0.19 $dB/km$, $\beta_2$ = ‚àí5.01 $ps^{2}km^{-1}$ and $\gamma$ = 1.2 $W^{‚àí1}km^{‚àí1}$) has been considered in the recirculating loop to best approximate
constant signal power evolution along the link. An EDFA is placed after the
fiber to compensate the span loss and ensure the same
launched power after each loop. A flat-top optical filter with a 3db bandwidth
of 1nm is used inside the loop to suppress the out-of-band (amplified
spontaneous emission) ASE noise. At the receiver, the signal is first aligned
in a particular polarization by a polarization controller  and then detected by an integrated coherent
receiver. The output electrical waveforms are sampled by a digital storage
scope (with a sampling rate of 80 GSa/s and a bandwidth of 33GHz) followed by off-line digital signal processing (DSP).
\begin{figure}
\centering
\includegraphics[width=8cm]{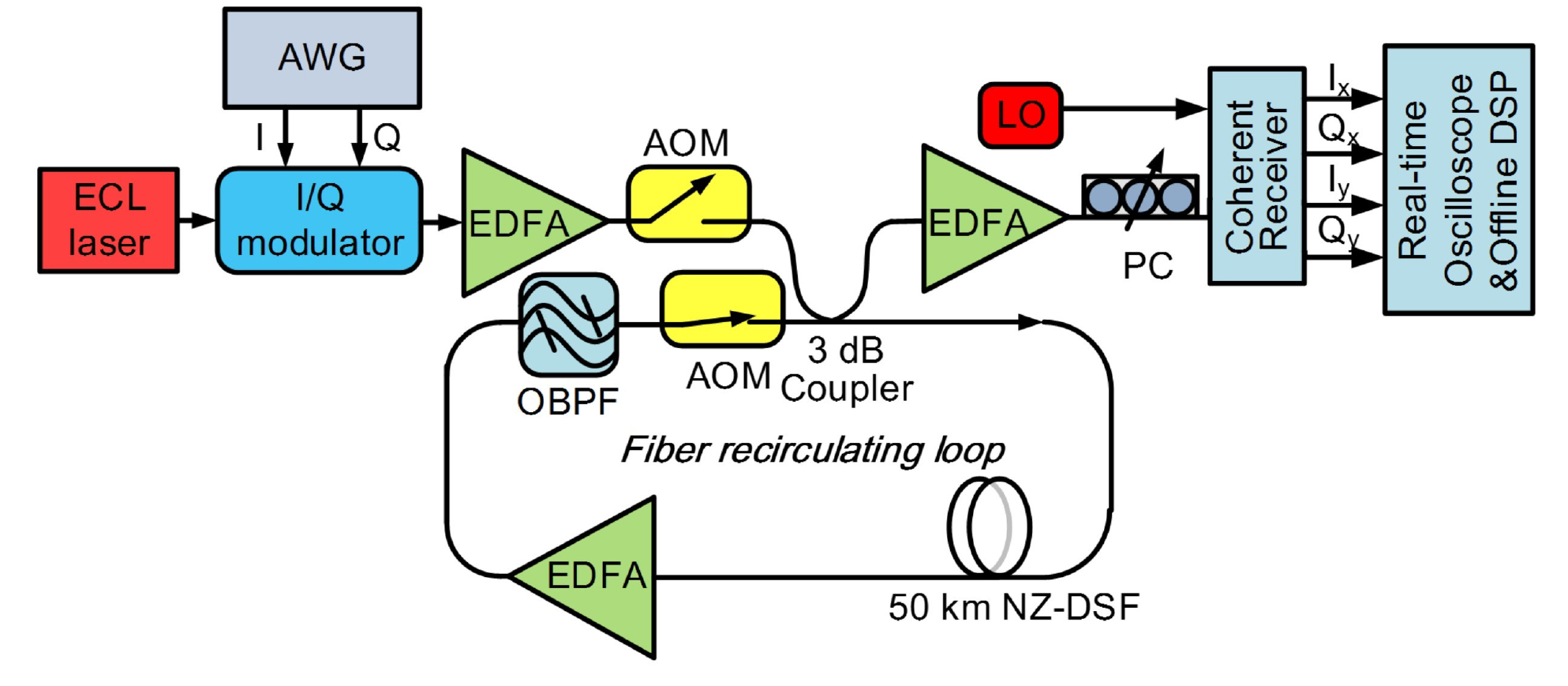}
\caption{
%Experimental setup. AWG: arbitrary waveform generator, AOM: acousto-optic modulator, OBPF: Optical band pass filter. PC: polarization controller.
Experimental setup. ECL: external cavity laser; AWG: arbitrary waveform generator; AOM: acousto-optic modulator; EDFA: erbium doped fiber amplifier; NZ-DSF: non zero dispersion shifted fiber; OBPF: Optical band pass filter; LO: local oscillator; PC: polarization controller.
}
\label{figExpSetup}
\end{figure}

% \begin{figure}
% \centering
% \subfloat[$A_{const}$]{\includegraphics[width=4cm]{FigNoiseTypeA}}\subfloat[$A_{}$]{\includegraphics[width=4cm]{FigNoiseTypeAt}}\\
% \subfloat[$C_{const}$]{\includegraphics[width=4cm]{FigNoiseTypeC}}\subfloat[$C_{}$]{\includegraphics[width=4cm]{FigNoiseTypeCt}}\\
% \subfloat[$B_{const}$]{\includegraphics[width=4cm]{FigNoiseTypeB}}\subfloat[$B_{}$]{\includegraphics[width=4cm]{FigNoiseTypeBt}}
% \caption{The effect of different types of noises on NFT eigen values.}
% \label{figNoiseType}
% \end{figure}

\begin{figure}
\centering
\subfloat[Experiment]{\includegraphics[width=4.5cm]{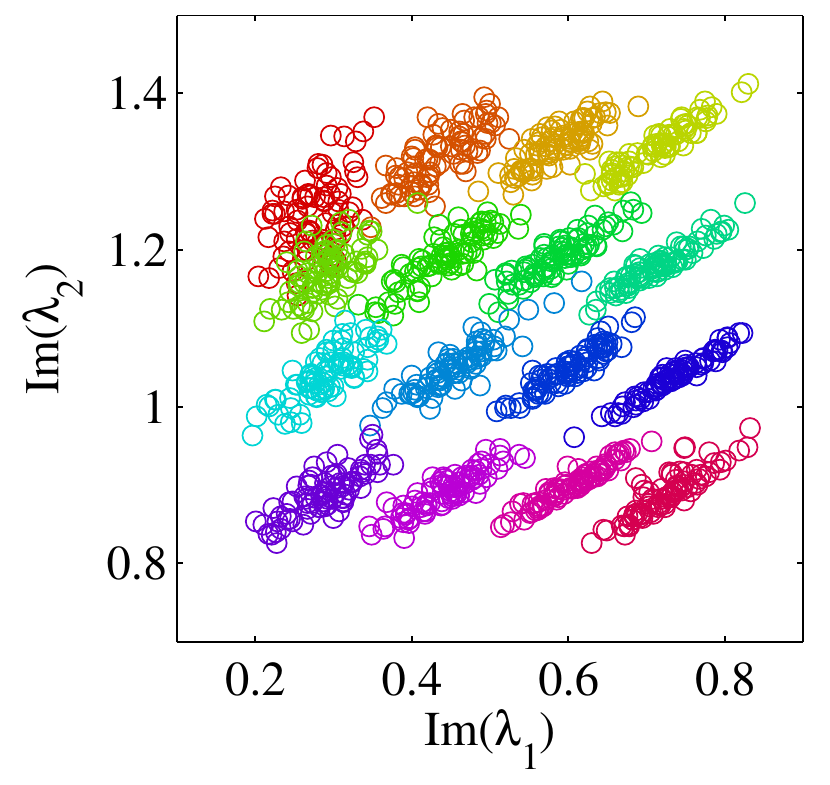}}
\subfloat[Simulated]{\includegraphics[width=4.5cm]{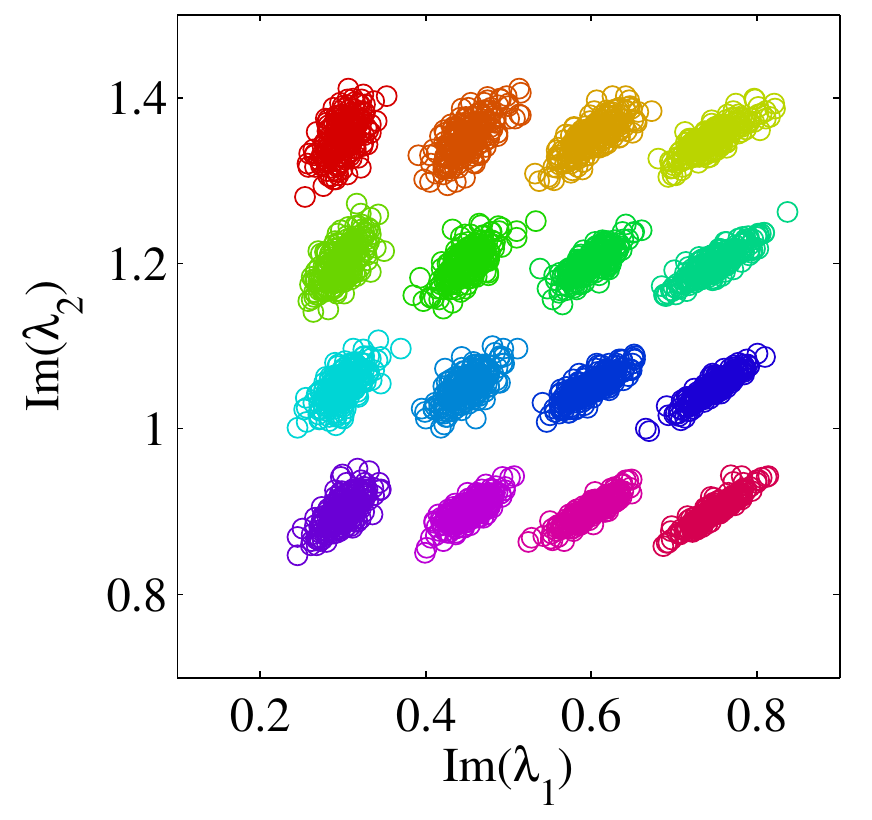}}\\
\subfloat[Experiment]{\includegraphics[width=4.5cm]{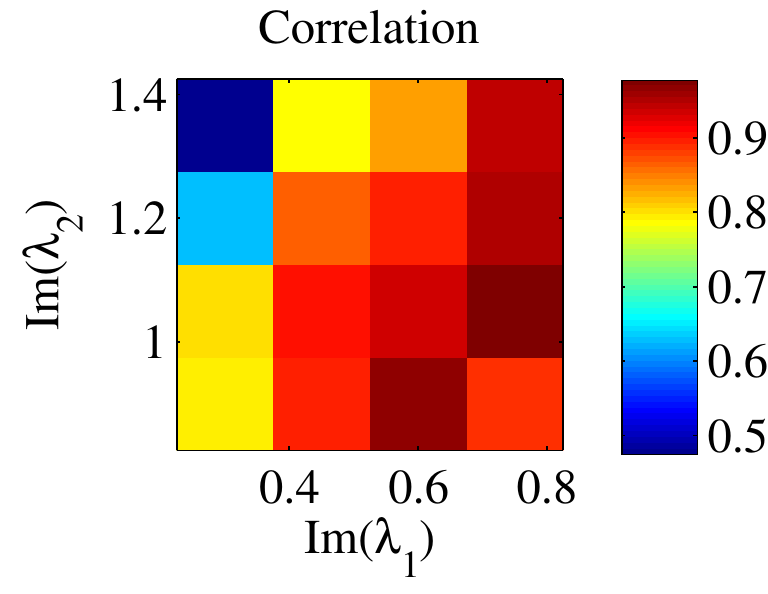}}
\subfloat[Simulated]{\includegraphics[width=4.5cm]{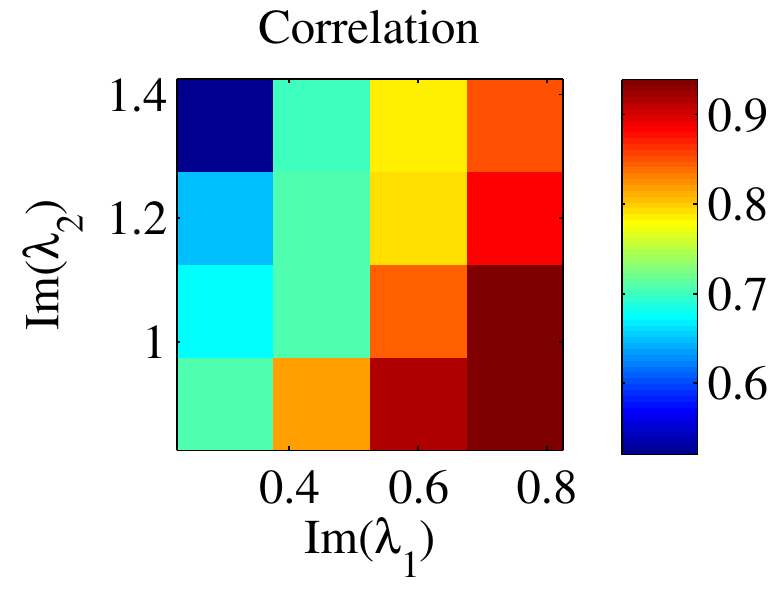}}\\
\caption{Experimental, (2a), and simulated, (2b), distribution of eigenvalues after $400$ km transmission. 2c and 2d, experimental and simulated contour plot of correlation parameter, respectively.}
\label{figExpResult}
\end{figure}

Figure \ref{figExpResult} shows the experimental and simulated received signal distributions of the group of 2-soliton pulses after propagating a distance of 400km, equivalent of 8 times circulation within the fiber loop. For each pulse set $(\lambda_{1},\lambda_{2})$ simulations were ran for 500 times, with noiseless input pulses but distributed random noise within the fiber. After propagation, the eigenvalues of these 500 outputs were then calculated using a forward difference method \cite{frank1}, see Simulation section in Methods \ref{Sim}. Figs \ref{figExpResult}a and \ref{figExpResult}b show the experimental and simulation distribution of each set of eigenvalues after propagation, respectively. A circular distribution of the two eigenvalues are expected for totally uncorrelated eigenvalues, however, the results show a linear-like distribution, which indicates a positive correlation. To quantify this we have calculated the sample correlation coefficient\footnote{
The sample correlation between $n$ samples of $(x_{i},y_{i})$ for $i=1, \ldots, n$ is defined as 
\[
\frac{\sum_{i=1}^{n}(x_{i} - \bar x)(y_{i} - \bar y) }{(n-1)s_{x} s_{y}}
\]
where 1) $\bar x$, $\bar y$ are sample means of $x$ and $y$, and  2) $s_{x}$ and $s_{y}$ are the sample standard deviations of $x$ and $y$.
}
 between $Im(\lambda_{1})$ and $Im(\lambda_{2})$ for each set of eigenvalues and represent them in Figs. 2c and d, for experimental and simulation results, respectively.   

%Figure \ref{fig_asecht} shows strong positive correlation between the two eigenvalues for all combinations. Note that for these results only noise associated with the propagation has been considered. The strong correlation allows better spectral efficiency in the perpendicular direction.  Strong positive correlations between the noises in the two eigenvalues can be seen in all experimental cases. The simulation results qualitatively agree with the experimental results in the following aspects:

Both experimental and simulation results show the following common characteristics that indicate the correlated nature of NFT eigenvalues:

(1) Both experiment and simulation show positive correlation represented by elongated elliptical shapes with a slightly different orientation of major axis.
From a signal design point of view, and as we demonstrate through a simulation below, one can appropriately leverage such correlation properties by more compact arranging of NFT eigenvalues in the direction of the minor axis of the elliptical distribution and achieve a higher spectral efficiency.

(2) In both cases, correlation decreases as $\lambda_2$ increases or $\lambda_1$ decreases.

However, some discrepancies between experimental and simulation results are also observed. Qualitatively, the orientation of the distributions of the two eigenvalues and their correlation are different for experiment and simulations results, see Fig. \ref{figNoiseType} (in supplementary materials). This could be explained by noting that in simulations, we only consider white Gaussian noise added during the propagation along the fiber. In the experiment, however, apart from the noise generated during propagation, noises are also induced during the pulse generation stage (e.g., when using the AWG to generate the electrical signals and the IQ modulator to generate optical signals.) and detection stages. In general, different types of noise can affect a signal at different stages of generation, propagation, and detection. In the supplementary materials, we have investigated collectively the effects of different types of noise on the correlation eigenvalue distribution.    

%%%%% Updated by Terence on 11/11/ 2016

Correlations of NFT eigenvalues has a significant impact on designing an NFT-based fiber optic network. In the following, we consider an example illustrating how to exploit such correlations to improve the performance of a system. 

\begin{figure}
\centering
\subfloat[]
{\includegraphics[width=4.5cm]{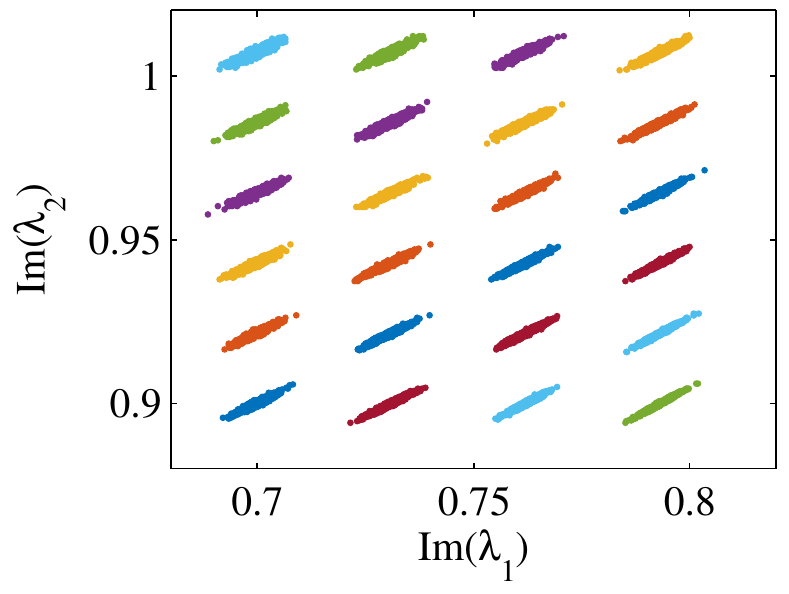}}
\subfloat[]
{\includegraphics[width=4.5cm]{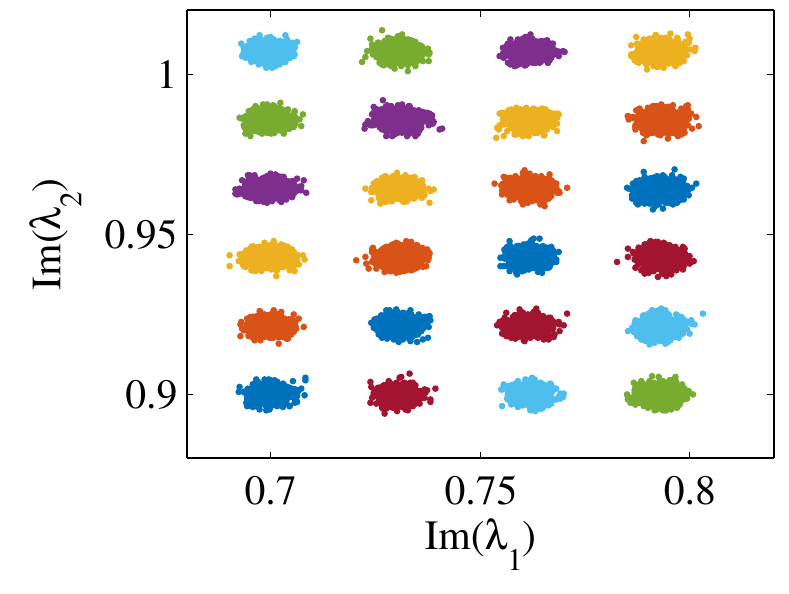}}\\
\subfloat[]
{\includegraphics[width=9cm]{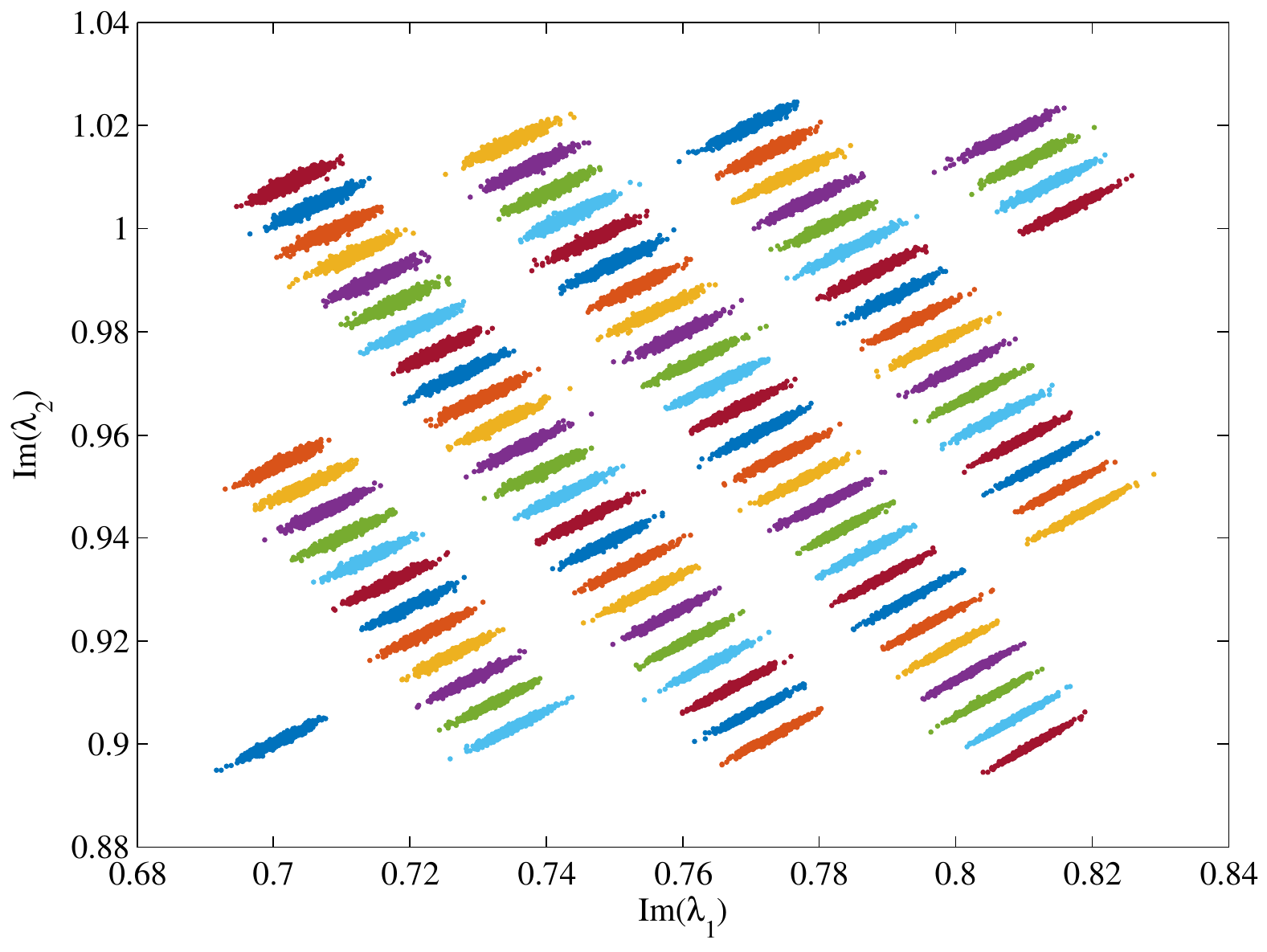}}
%
%\caption{Distribution of eigenvalues for constellation of 2-soliton pulses after propagating for 0.1 normalized distance. (a) distribution of eigenvalues for a $4\times6=24$ constellation with eigenvalue spacing of $(0.04,0.02)$. (b) distribution of eigenvalues for a $4\times6$ constellation but assuming no correlation. (c) distribution of eigenvalues for a $83$ constellation with eigenvalue spacing of $(0.??,0.???)$.}
\caption{Distribution of eigenvalues for constellation of 2-soliton pulses after propagating for 0.1 normalized distance. (a) distribution of eigenvalues for a $4\times6=24$ constellation.  (b) distribution of eigenvalues for a $4\times6$ constellation but assuming no correlation. (c) distribution of eigenvalues for a $83$ constellation.}
\label{figcorr1}
\end{figure}

{\bf Example:}
Consider input signal constellation of 2-solitons pulses, with eigenvalues 
$\lambda_{1}  \in  [0.7 \imag , 0.74 \imag , 0.78 \imag , 0.82 \imag]   $ and $\lambda_{2}\in  [0.9 \imag , 0.92 \imag , ..., 1.00 \imag ,1.02 \imag]$. For each input, we propagate the signal in a fibre for a normalised distance of 0.1.  Figure \ref{figcorr1}(a), shows the distribution of the two eigenvalues for the output pulses, clearly showing correlation between the eigenvalues. 
Now, suppose the correlation is ignored. Then the decoder will assume in the worst case that the two eigenvalues are independently distributed. Figure \ref{figcorr1}(b) shows the distribution of two eigenvalues assuming that they are in fact independently distributed. This leads to the requirement that the eigenvalues of input signals should be separated sufficiently enough to ensure that the decode can distinguish different inputs with high accuracy. 
On the other hand, by exploiting correlations, we can pack more inputs with eigenvalues in range of $[0.7 \imag , 0.82 \imag]  $ and $[0.9 \imag , 1.02 \imag]$ such that the decoder can still decode with low error probability. Figure, \ref{figcorr1}(c) shows a 10 fold larger constellation, using the same range of eigenvalues for $\lambda_{1}$ and $\lambda_{2}$, but denser packing of the eigenvalues of input signals. In this example, the data rate has been increased from  $\log 24$  bits per symbol to  $\log 84$  bits per symbol, equivalent to  approximately 39 percent increase in data rate. Clearly, the actual increase in data rate will depend on a lot of other factors, such as power and bandwidth constraint. 
% \end{example}

{\bf Remark: }
The above example is used only to demonstrate how correlations can be exploited to increase data rate.

\begin{figure}
\centering
\includegraphics[width=9cm]{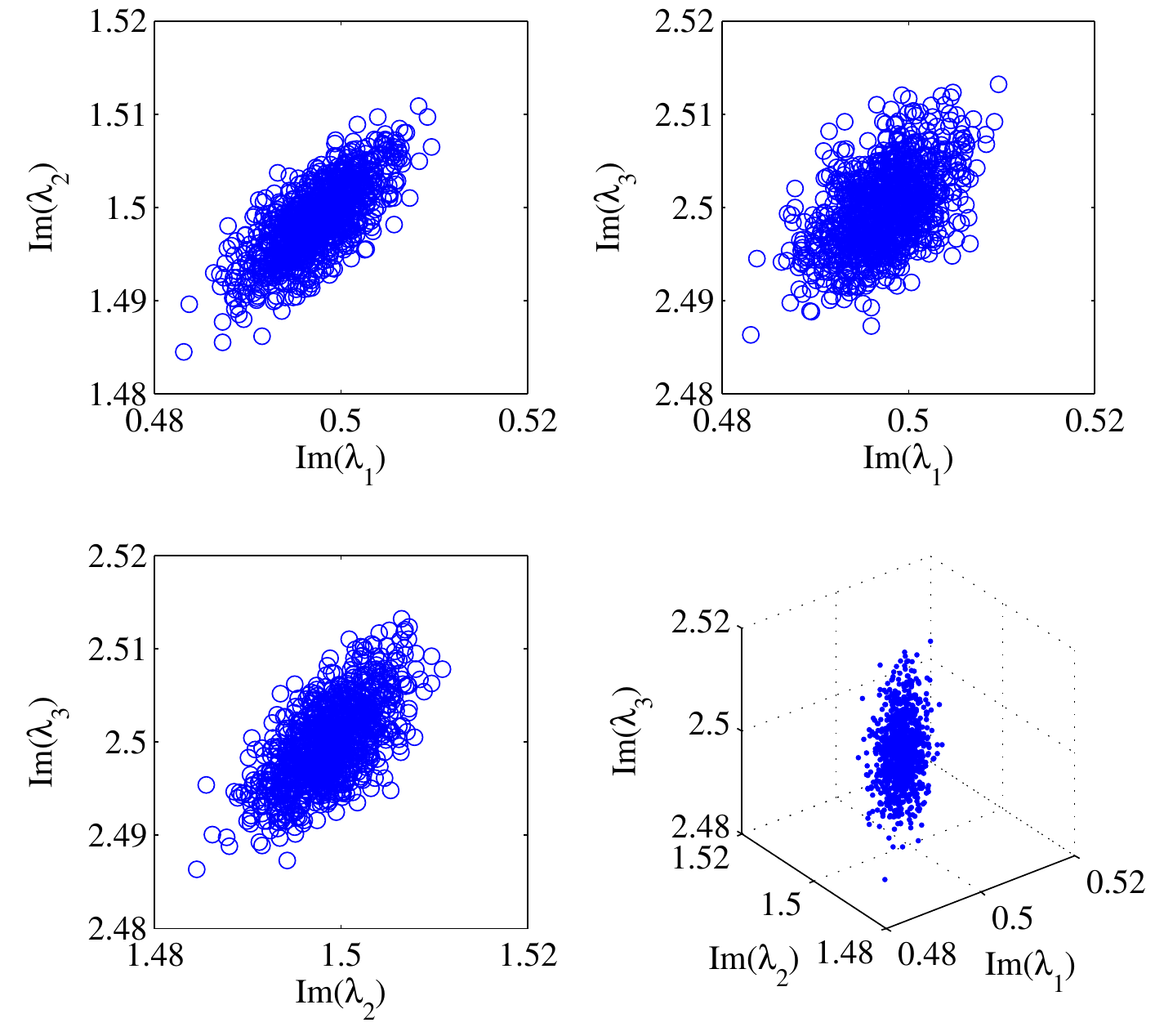}
\caption{(a), (b), (c) Correlations between all possible pairs of eigenvalues for a 3-eigenvalue input signal. (d) correlation of all eigenvalues in 3 dimensional parameter space.}
\label{fig_3eigsCorr}
\end{figure}

\bigskip
Further simulation studies reveals that the correlation exists in systems with larger number of discrete eigenvalues, e.g. 3. Figure \ref{fig_3eigsCorr} shows a case with 3 discrete eigenvalues $ 0.5 \imag, 1.5 \imag $  and $2.5 \imag$. The subplots of Fig. \ref{fig_3eigsCorr} show the correlation between eigenvalues $\lambda_1$ and $\lambda_2$, $\lambda_1$ and $\lambda_3$, $\lambda_2$ and $\lambda_3$ as well as a distribution of the eigenvalues in a 3-D parameter space. Similar correlations can also be found in systems with even more eigenvalues.

%%%%%%%%%%%%%%%%%%%%%%%%%%%%%%%
%% Section IV 
%%%%%%%%%%%%%%%%%%%%%%%%%%%%%%%

\def\lambdav{{\Lambda}}

\section{{\bf Results:} Modeling eigenvalues perturbation}

One fundamental challenge in designing eigenvalue communications system is to characterize the noise in the discrete eigenvalues. 
In the previous section, we have discussed the correlation of discrete eigenvalues for short distance (of normalized length 0.1)  experimentally and numerically. In the following, we will  develop a full model of eigenvalue perturbation, based on our two observations on properties of NFT discrete eigenvalue perturbation in an optical communication system:
\begin{enumerate}
\item  {\bf Split Step Method for noise;} deterministic (due to nonlinearity and dispersion) and stochastic (e.g., due to all signal amplification) noise effects can be separated,  in a similar fashion as linear and nonlinear processes in Split Step Method, and 

\item 
{\bf No noise hysteresis;}, the effect of noise associated with each segment of optical communication can be dealt independently. 
\end{enumerate}

% better undrestanding of In the previous section, we have 
% investigated eigenvalue perturbation via simulations and experiments. In the following, we will address the problem via analysis. 

First, we will state the framework based on which the eigenvalue noise perturbation model is developed. When a signal propagates along a fibre, it will be distorted by various channel impairments\footnote{In this paper, we will assume that the fibre loss can be perfectly compensated by inline distributed amplification.}  such as noises and fibre nonlinearity and dispersion,  causing its shape to change during propagation. To model the process, we treat a fibre  as a concatenation of many short fibre segments. 
% This view is not new. For example, the split step Fourier method is also based on a similar idea. 
%
As each segment is short,   fibre nonlinearity and noise can often be modeled separately. Under this model, when a signal propagates in a segment, it will undergo two phases. In the first phase, it will be distorted by fibre nonlinearity, assuming  that the segment is noiseless. Then in the next phase, a white Gaussian noise (whose power is proportional to the length of the fibre segment) will be added. The resulting signal will then become the input of the next segment, and the same process will continue until it reaches  the end of the fibre, see Fig. \ref{figsplit}.

Modeling a fibre as a concatenation of short segments is only the first step. 
Due to fibre nonlinearities and the coupling effects between noises and signals, it is still very difficult to derive  an analytic model to characterise the eigenvalues perturbations. 
%The idea is not totally new, very much analogous to the split step Fourier method. 
 In the following, we aim to simplify the model.

%%%%%%%%%%%%%%%%%%%%%%

\def\error{\epsilon} 
 
\subsection{Simplification 1: Noise Decoupling} 

\begin{table}[]
\centering
\caption{Notations}
\label{my-label}
\begin{tabular}{|c| l | l l l}
\cline{1-2} \cline{4-5}
\multicolumn{1}{|c|}{Symbol} & \multicolumn{1}{c|}{Definition}                                                                                                    & \multicolumn{1}{l|}{} & \multicolumn{1}{c|}{Symbol}           & \multicolumn{1}{c|}{Definition}                                                                       \\ \cline{1-2} \cline{4-5} 
$q_m(t)$                        & \begin{tabular}[c]{@{}l@{}}output of signal after \\ propagating $m$ segment\end{tabular}                                          & \multicolumn{1}{l|}{} & \multicolumn{1}{l|}{$\lambdav_m$}         & \multicolumn{1}{l|}{\begin{tabular}[c]{@{}l@{}}discrete eigenvalues\\  of $q_m(t)$\end{tabular}}      \\ \cline{1-2} \cline{4-5} 
$\bar q_m(t)$                   & \begin{tabular}[c]{@{}l@{}}output of signal after \\ propagating $m$ segment,\\ assuming no added \\stochastic noises\end{tabular} & \multicolumn{1}{l|}{} & \multicolumn{1}{l|}{$\hat \lambdav_m $} & \multicolumn{1}{l|}{\begin{tabular}[c]{@{}l@{}}discrete eigenvalues  \\of $\hat q_m(t)$\end{tabular}} \\ \cline{1-2} \cline{4-5} 
$n_m(t)$                        & \begin{tabular}[c]{@{}l@{}} noises added in the \\ $m$ segment\end{tabular}                                                & \multicolumn{1}{l|}{} & \multicolumn{1}{l|}{$g(\lambdav)$}        & \multicolumn{1}{l|}{a function of $\lambdav$}                                                          \\ \cline{1-2} \cline{4-5}
$\hat q_m(t)$                   & $\bar q_m(t) + n_m(t)$                                                                                                             &                       &                                          &                                                                                                       \\ \cline{1-2}
\end{tabular}
\end{table}

%While the signal is distorted by fibre nonlinearity (in the first phase) and noise addition (in the second phase), it is worthy to point out that the discrete eigenvalues of the signal is in fact  unaffected by the fibre nonlineariy. In other words, after the first phase, the output of the segments and the input of the segments have the same discrete eigenvalues. As such,  perturbation in eigenvalues is only due to noise addition. 

One of the challenges in deriving a model for characterising the perturbation of discrete eigenvalues is due to the coupling effects of stochastic signal dependent noises (e.g., due to inline amplification) and the distortion due to nonlinearities and dispersions. In the following, we propose a new simplification paradigm to decouple the two noise effects.
Our proposed simplification is based on the observation that  \emph{perturbation of eigenvalues at the end of the fibre can be accurately modelled by ``summing up'' all the small perturbations of eigenvalues in the segments}.

Suppose the input to the fibre is $q_{0}(t)$. Divide the fibre into $M$ segments and let $q_{m}(t)$ be the output of the signal after propagating   $m $ segments (or equivalently the input to the $(m+1)^{th}$ segment). Define 
$\lambdav_{m} $  as the set of discrete eigenvalues of $q_{m}(t)$. 
Let $g(\lambdav_{m})$ be a (scalar or vector valued) function of $\lambdav_{m}$ of interest.
% The following are some examples:
% \begin{align}
% g(\lambdav_{0}) \triangleq \min_{i \in \Lambda} \text{imag}(\lambdav^{(i)}_{0}), 
% \end{align}
% \begin{align}
% g(\lambdav_{0}) \triangleq \max_{i \in \Lambda} \text{imag}(\lambdav^{(i)}_{0}), 
% \end{align}
% or
% \begin{align}\label{eq25}
% g(\lambdav_{0}) \triangleq \sum_{i \in \Lambda} \text{imag}(\lambdav^{(i)}_{0}). 
% \end{align}

\begin{figure}[h]
\centering
 \includegraphics[width=9cm]{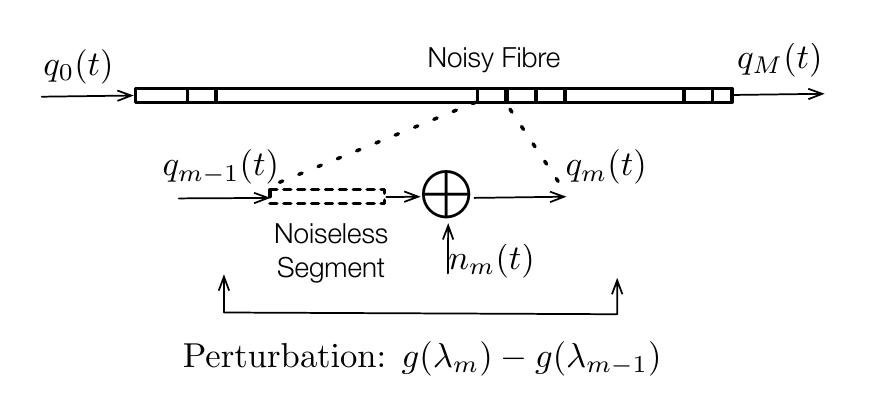}
\caption{Channel Model: Fiber is considered as a concatenation of $M$ segments. The perturbation of eigenvalues (or their function values) is modeled as the accumulation of many perturbations caused by the addition of noises in each segment. Each perturbation $\error_{m}$ is further modelled as independent, depending  only on the deterministically distorted signal $\bar q_{m}(t)$.t}
\label{figsplit}
\end{figure}

Notice that 
\begin{align*} 
g(\lambdav_{M} ) - g(\lambdav_{0})   = \sum_{m=1}^{M}  g(\lambdav_{m}) - g(\lambdav_{m-1}) . 
\end{align*}
Therefore, we can characterise the perturbation $g(\lambdav_{M}) - g(\lambdav_{0}) $ by characterising the perturbation $ g(\lambdav_{m}) - g(\lambdav_{m-1}) $ for all $m$, see Fig. \ref{figsplit}.

Observing each individual term, the perturbation $g(\lambdav_{m}) - g(\lambdav_{m-1}) $ is caused by the injection of a noise at the $m^{th}$ segment. 
First, we want to point out that  the perturbation  depends on   both  the injected noise $n_{m}(t)$ and the input of the segment $q_{m-1}(t)$ (which in turn also depends on  the noises added in the previous segments).  However, we claim that; \textit{the influence due to the noise added in the previous segments are insignificant (and hence can be ignored)}.

Let $\bar{q}_{m}(t)$ be the output of the signal after propagating through $m^{th}$ segments, assuming there are no noises. Hence,  $\bar{q}_{m}(t)$ is a deterministic signal. Due to fibre nonlinearity, its shape will vary with $m$. However, the discrete eigenvalues for all $\bar{q}_{m}(t)$ remained unchanged. Let  $n_{m}(t)$ be the noise  added in the $m$ segments, 
\begin{align}\label{eq24}
\hat{q}_{m} ( t)  = \bar{q}_{m} ( t) + n_{m}(t)
\end{align}
and 
$\hat{\lambdav}_{m}$ be its set of discrete eigenvalues. In other words, $\hat{q}_{m}$ is obtained by propagating $q_{0}(t)$ noiselessly across $m$ segments, followed by the addition of the noise $n_{m}(t)$. See Figure \ref{figsplit}. 
%While discrete eigenvalues of $\bar{q}_{m} ( t) $ are the same for all $m$,   the perturbation of the eigenvalues (i.e., $\hat{\lambdav}_{m}$) induced by the  addition of the same noise will depend on specific $\bar{q}_{m} ( t)$. 
We claim that $g(\hat{\lambdav}_{m}) - g(\lambdav_{0})$ is in indeed a good approximation for $g(\lambdav_{m}) - g(\lambdav_{m-1})$.
%
%
%
%
%Specifically, we propose the following approximation  
%\begin{align}
%g(\lambdav_{m}) - g(\lambdav_{m-1})  & \approx g(\hat\lambdav_{m}) - g(\bar\lambdav_{m-1})  \\
% & = g(\hat\lambdav_{m}) - g(\lambdav_{0}). 
%\end{align}
Let 
$
\error_{m} = g(\hat\lambdav_{m}) - g(\lambdav_{0})
$.  Then 
$
g(\lambdav_{M}) - g(\lambdav_{0})   \approx \sum_{m=1}^{M}  \error_{m} 
$
or equivalently, 
\begin{align} 
g(\lambdav_{M})    \approx  g(\lambdav_{0}) +  \sum_{m=1}^{M}  \error_{m}. \label{eq27}
\end{align}

\subsubsection*{Benefits}
The above approximation points out that the end-to-end perturbation is now modelled as the sum of a collection of independently distributed local perturbations (as  $\bar{q}_{m}$ is deterministic and the noise $n_{m}(t)$ is  independently distributed for all $m$). The main benefit of the model is its simplicity, decoupling the stochastic noises (caused by inline amplification) from the deterministic dispersion and nonlinearities. 

To be more precise, we have already seen that the local perturbation of the eigenvalues
\[
g(\hat \lambdav_{m}) - g(\Lambda_{0})
\] 
depends on $m$ (and more precisely  the signal that enters the $m$ fibre segment, i.e., $\bar q_{m}(t)$). 
This same argument also applies to the local perturbation that
\[
g(\lambdav_{m}) - g(\lambdav_{m-1})
\] 
also depends on the signal $q_{m-1}(t)$ which is stochastic in nature due to the noises added in the previous $m-1$ segments. In that case, the stochastic random noise and the deterministic dispersion and nonlinearities will couple with each other. Furthermore, 
\[
g(\lambdav_{m}) - g(\lambdav_{m-1})
\] 
will become correlated for different $m$. 
Our approximation decouples the two effects, resulting in a simpler mode. 
%\[
%g(\lambdav_{m}) - g(\lambdav_{m-1}) \approx g(\hat \lambdav_{m}) - g(\Lambda_{0}).
%\] 
Through the approximation, we also break the correlation among local perturbations, making channel analysis  more manageable.

%The above approximation means that 
%\begin{enumerate}
%\item
%the perturbation error $g(\lambdav_{M}) - g(\lambdav_{0})$ is modeled/approximated as  
%the sum of errors  $\error_{m}$; 
%\item 
% each individual perturbation $\error_{m}$  is also independently distributed -- .   
%
%\end{enumerate}

%\begin{figure}[h]
%\centering
%\includegraphics[width=8cm]{split2}
%\caption{Approximated Channel Model: perturbation in each segment is independent of noise in previous section and only depends on deterministic signal at the beginning of the segment.\textit{\textbf{ ???? suggest to put figure 6 and 7 together and specify that q-bar in fig7 is deterministic input function to the segment???? }}}
%\label{figsplit2}
%\end{figure}

\subsection{Validation}
% In the previous subsection, we  offered a simple way to model eigenvalue perturbation as a sum of independent smaller errors. 

In the following,  we will validate Eq. (\ref{eq27}) through numerical simulation and experiment. We investigate the perturbation of discrete eigenvalues of a 2-soliton input signal, $q_{0}(t) = 2 sech(t)$, which has two discrete eigenvalues at $0.5 \imag$ and $1.5 \imag$. Specifically, the function $g(\cdot)$ is a vector valued function, corresponding  to the imaginary parts of the two discrete eigenvalues of the signals propagating along the fibre. 
First, using numerical simulation (details of simulation method are given in \ref{Sim}), we illustrate  that the perturbation $\error_m = g(\hat{\lambdav}_{m}) - g(\lambdav_{0})$  is signal dependent (i.e., depending on $\bar{q}_{m}(t)$).

% \tc{ (SHALL WE add a series of figures illustrate the shape of $\bar q_{m}(t)$? }

\begin{figure*}
\centering
 \includegraphics[width=18cm]{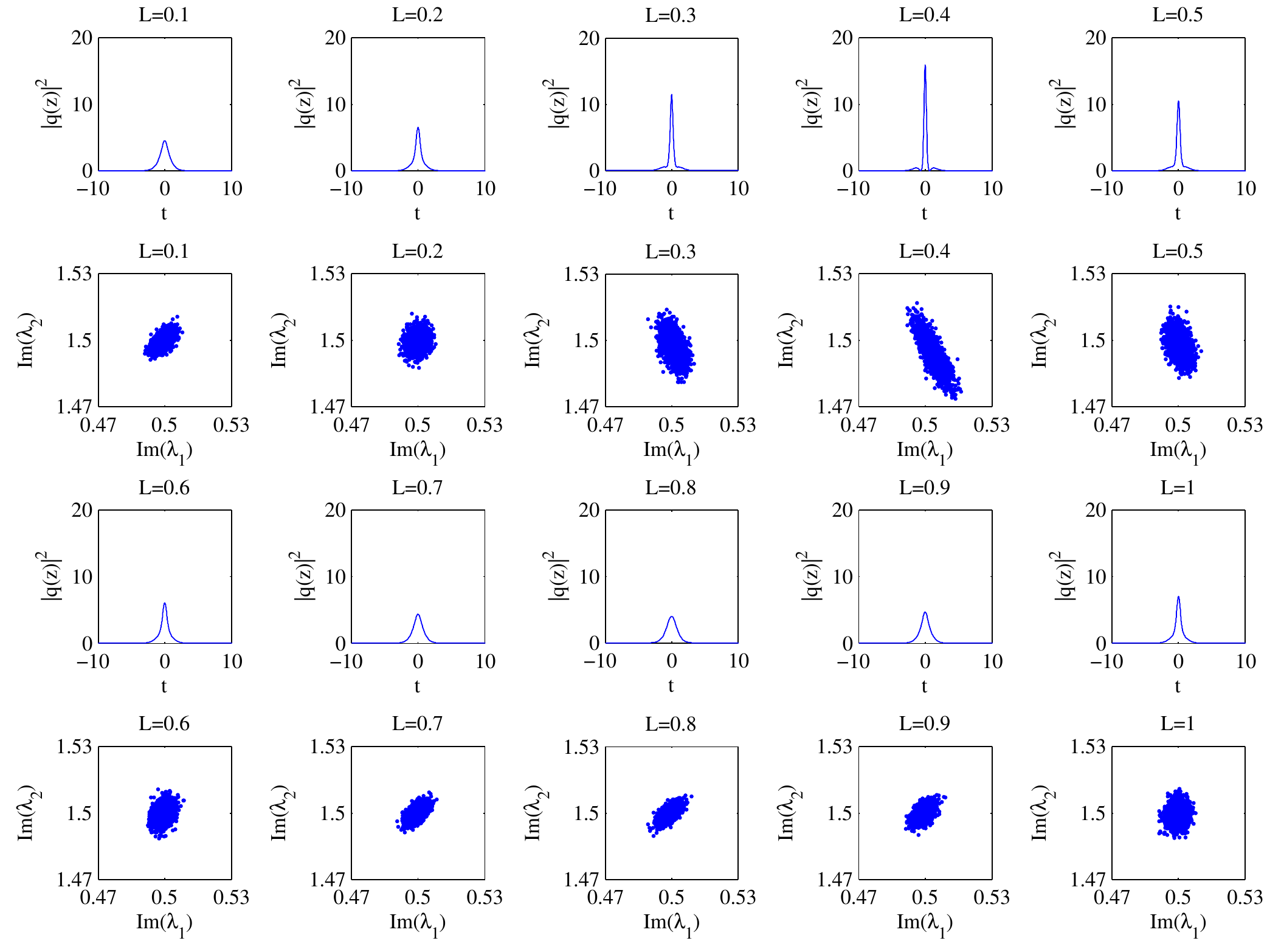}
\caption{Scattering plot for  $\hat{\lambdav}_{m}$ at different normalized propagation lengths: 0.1, 0.2, 0.3, ...,0.9,1.0.}
\label{addition}
\end{figure*}

We plot the ensemble of two discrete eigenvalues  of $\hat{\lambdav}_{m}$ for various $m$. Since the time domain signal will change its shape as it propagates,  our numerical example clearly shows that not only the two eigenvalues are all correlated, but how they correlate depend on $m$ as well (See Figure \ref{addition}). This supports our observation that the statistics of  $g(\hat{\lambdav}_{m}) $ (and hence also   $g(\hat{\lambdav}_{m}) - g(\lambdav_{0})$  as $g(\lambdav_{0})$ is a constant)  depends on 
$\bar{q}_{m}(t)$. It also suggests that the statistics of $ g(\lambdav_{m}) - g(\lambdav_{m-1}) $  will also depend on $q_{m-1}(t)$ which, strictly speaking, will also be influenced by the noises injected in the previous segments.

Observing the scattering plots for $\hat{\lambdav}_{m}$, the mean and covariance matrix of  $\hat{\lambdav}_{m}$ depends on $m$ (and more precisely, on $q_{m-1}(t)$). Clearly, it also depends on the noise power. In order to better understand the influence of
$q_{m-1}(t)$ on $\hat{\lambdav}_{m}$, we will focus on the eigenvectors (in particular the principal one) of the covariance matrices. %It can be argued that the eigenvector is insensitive to the noise power, under our hidden linearity assumption.  

%
% (\textbf{??? Instead of b we may consider Q in eq. 19, we get the same phase ?????})
% 
 
Now, consider a 2 soliton $q_{0}(t) = 2 sech(t)$, which has two discrete eigenvalues at $0.5 \imag$ and $1.5 \imag$. 
Then $\bar{q}_{m}(t)$, obtained by propagating $q_{0}(t)$ for $m$ segments, are also 2-solitons, with discrete eigenvalues at $0.5 \imag$ and $1.5 \imag$. Let $Q^{(d)}_{m}(0.5 \imag) $ and $Q^{(d)}_{m}(1.5 \imag)$  be its corresponding spectral amplitudes. Then they will vary as $m$ increases (i.e., as the signal propagates). 
%
%
% The values of the  spectral amplitudes  $Q^{(d)}(0.5 \imag) $ and $Q^{(d)}(1.5 \imag)$ for $\bar{q}_{m}(t)$ will evolve with $m$ (or equivalently the normalised distance the signal has transversed). 
% 
 To make it more precise, if the normalised distance of each fibre segment is  
$\Delta_{L}$, then $q_{m}(t)$ is the signal after propagating $q_{0}(t)$ for a distance of $m \Delta_{L}$.  If we define $\theta(m \Delta_{L}) $ as the ratio
$
Q^{(d)}_{m}(0.5 \imag) / Q^{(d)}_{m}(1.5 \imag), 
$
  then   
\[
\theta(m \Delta_{L}) =  e^{ 4  m \Delta_{L} ( \lambda_{1}^{2} -  \lambda_{2}^{2})  \imag }
\]
where $\lambda_{1} = 0.5 \imag $ and $\lambda_{2} = 1.5 \imag$.
We call $4 m \Delta_{L}( \lambda_{1}^{2} -  \lambda_{2}^{2}) $ the \emph{nonlinear phase difference}.

\begin{figure}
\centering
\includegraphics[width=9cm]{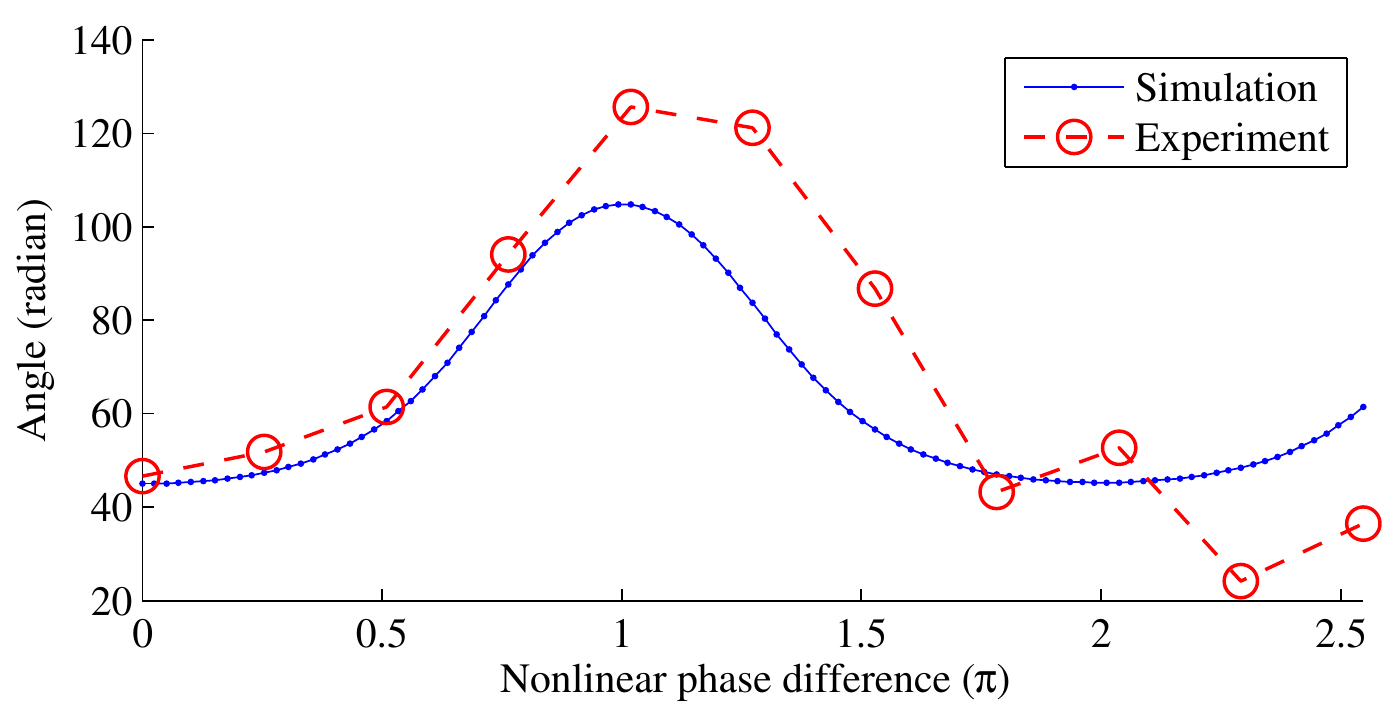}
\caption{NFT phase difference between $\lambdav_1$ and $\lambdav_2$}
\label{SimExpAngle}
\end{figure}

In our example, we focus on the imaginary parts of the two discrete eigenvalues. Figure \ref{addition} plots the two discrete eigenvalues, which shows that they are correlated with each other. Also, the correlation is different for different $m$. For each scattering plot, we can estimate the covariance matrix between the imaginary parts of the two discrete eigenvalues. We can also plot the  angle of the principal eigenvector\footnote{
The principal eigenvector of a covariance matrix is the matrix's eigenvector (often of unit length) with respect to the largest eigenvalue. In the case of a $2\times 2$ matrix, we can further represent the vector by the angle it makes with the horizontal axis. 
} of the covariance matrix.  
It turns out that the principal eigenvector  of the covariance matrix for $\hat{\lambdav}_{m}$ depends only on the nonlinear phase.  
%\tc{(modulo $2 \pi$), needs to be changed if we use a different notations} 
%
Simulation results are shown in  Figure \ref{SimExpAngle}, which  clearly indicate that the principal eigenvector, and also the covariance matrix for $\hat{\lambdav}_{m}$ are different for different $m$. Note that in the simulation, a noise is added to $\bar{q}_{m}(t)$, which is a result of propagating $q_{0}(t)$ noiselessly for a distance of $m \Delta_{L}$.
   
To verify our observation and to support the simulation result, we also consider the following experiment. Our experiment set up is very similar to the one in Figure \ref{figExpSetup}. EDFAs are used to amplify signals to combat signal loss. We consider a range of propagation distance up to 1500 km which consisting of 30 loops, where every 3 loops correspond to a 0.1 normalized length. This is to mimicking the generation of $\bar q_{m}(t)$.  Figure \ref{SimExpAngle} shows the experimental values for the orientation of principal eigenvector of covariance matrix (red circles). A qualitative agreement is observed between experiment and simulation results. This supports the assumption that eigenvalue distribution depends on signal at that point.

Note that, in experiments, noises are added at the transmitter, the receiver and also during propagation. However,  the propagation noises is small (compared to the other noises), since the propagation distances are all small in our experiments. In addition,  the transmitter noise is the same in all the experiments. Therefore, the difference in the distribution of eigenvalues observed at the output for different propagation lengths are only due to the receiver noise (See supplementary material attached). In our experiment, the receiver noise acts as the role of the point noise injected in the simulation. Careful examination of the simulation and experiment results in Figure \ref{SimExpAngle}, shows a good agreement when the propagation distance is small. However, the two results start to become less agreeable for longer distances. This can also be explained by the fact that the propagation noise increases for longer propagation distances.

%%%%%%%%%%%%%%%%%%%%%%%%%%%
So far,  we have demonstrated that the local perturbation caused by the injection of noise in a segment is not identically distributed and depends on pulse shape of the input at each segment. Next, we want to show that the overall perturbation $g(\lambdav_{M}) - g(\lambdav_{0})$ can be approximated by accumulating individual smaller perturbations $\error_{m}$. In other words, we want to show that the approximation \eqref{eq27} is indeed fairly accurate.

\begin{figure}
\centering
 \includegraphics[width=9cm]{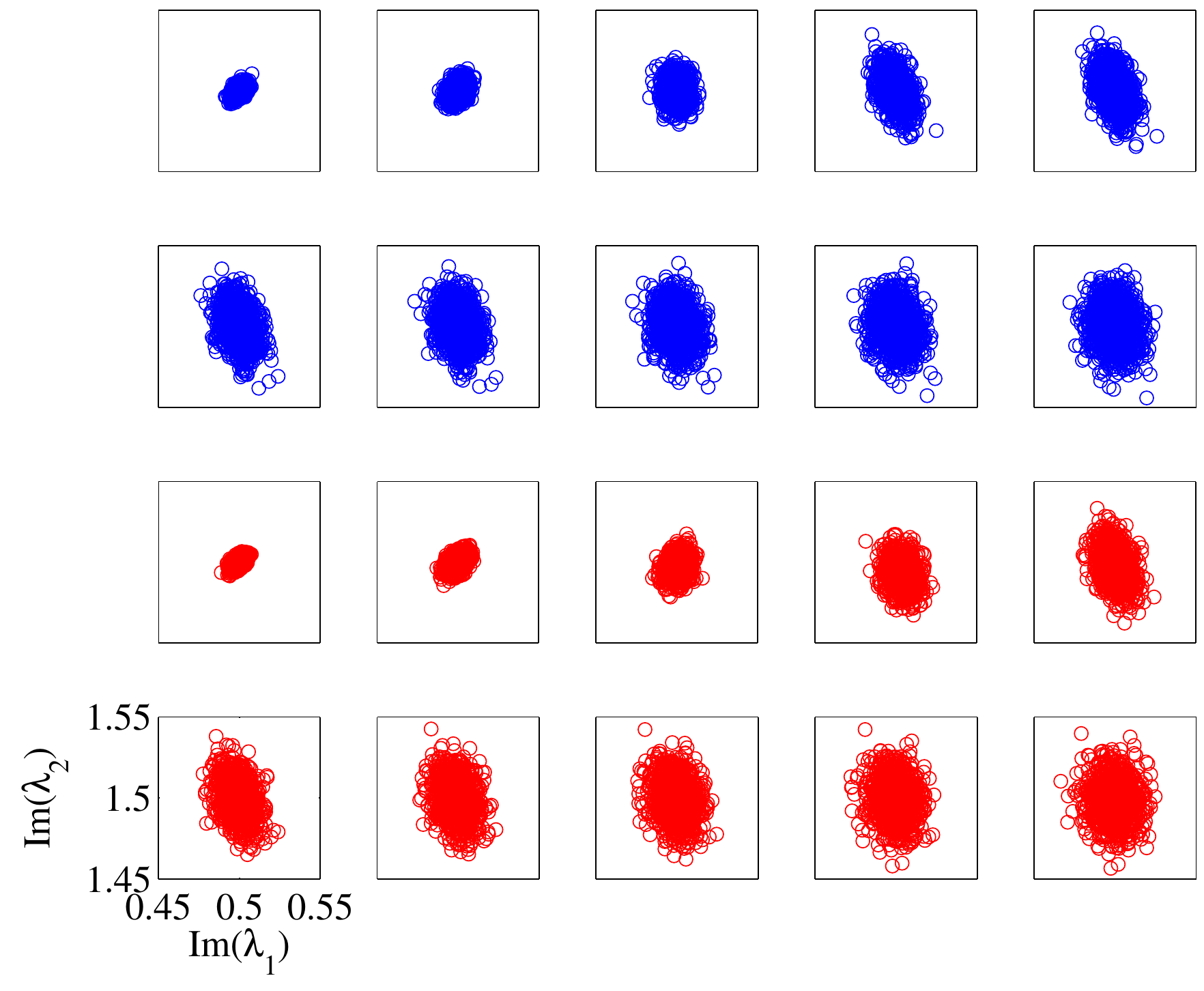}
\caption{Comparison of the accumulated perturbations with (top blue) and without (bottom red) the approximation Eq.~\eqref{eq27} for $m$ values ranging from 0.1 to 1 with a step of 0.1.}
\label{FigNoiseAngle}
\end{figure}

To validate, we will consider  the following numerical example. 
We consider the special case when $g_{0}(t)$ is a 2-soliton. In particular, we are  interested in the imaginary parts of the two discrete eigenvalues  of the input and output signals. In other words
\[
g(\Lambda) = [Im(\lambda_{1}), Im(\lambda_{2})]^{\top}
\]
where 
$\Lambda = (\lambda_{1},\lambda_{2})$.
Here, we consider various choices of $m$ (corresponding to propagation distance from 0.1 to 1).
Results are shown in Figure \ref{FigNoiseAngle}.

Let $\lambdav_{m} \triangleq (\lambda_{m,1} , \lambda_{m,2})$ be the discrete eigenvalues of $q_{m}(t)$, the signal obtained by propagating the input signal across $m$ fibre segments. 
The lower scatter plot (formed by the red circles) in  Figure \ref{FigNoiseAngle} is obtained by  plotting 
$Im(\lambda_{m,1} )$ against $Im(\lambda_{m,2} )$ for  various  $m$. 
 The  upper scatter plot (formed by the blue circles) in  Figure \ref{FigNoiseAngle} is obtained using the approximation \eqref{eq27}, defined as the sum of $g(\Lambda_{0})$ and a set of local perturbations caused by the addition of noises added throughout each segment. Specifically, the RHS of \eqref{eq27} is 
\[
\left[
\begin{array}{c}
0.5 + Im \left( \sum_{\ell  =1 }^{m} (\hat{\lambda}_{\ell, 1}  - 0.5 \imag) \right) \\
1.5 +  Im \left( \sum_{\ell  =1 }^{m} (\hat{\lambda}_{\ell,2}  - 1.5 \imag) \right) 
\end{array}
\right]
\]
where $\hat\Lambda_{\ell} = (\hat\lambda_{\ell,1}, \hat\lambda_{\ell,2})$ is the discrete eigenvalues of $\hat q_{\ell}(t)$ obtained by propagating the input signal $q_{0}(t)$ for $\ell$ segments followed by the addition of a local noise $n_{\ell}(t)$. Now, if we compare the two sets of scatter plots, 
we can see immediately that the two plots look extremely similar. This indicates that one can model the eigenvalues perturbation for $\lambdav_{m}$ pretty accurately, by using our approximation.

\subsection{Simplification 2: Noise decomposition}\label{sec:noisedecomposition}

In the previous section, we proposed  a simple  model to characterise the eigenvalues perturbation by modelling separately the perturbation in each individual segment. 
%In the following, we further analyse how to  simplify the noise addition model in each segment. 
The  noise added in each segment can be decomposed as the sum of many independent ``noise components". Depending on the decomposition, a noise component can be the noises added to a specific narrow frequency band or a time-interval.   In our earlier work \cite{Kashif2015}, we have  demonstrated  that noises added in frequency bands outside the signal frequency band will  have minimal impacts on the perturbation of eigenvalues. In other words, in the context of detecting the eigenvalues, out of band noises are essentially relevant. Now, the natural question thus is:  \emph{Which ``noise component'' will contribute the most to the perturbation of eigenvalues?} 
A complete answer to the question remains unknown. In this paper, we will focus on specific noise components  and investigate its contributions to eigenvalue perturbation. 

We are interested in the noises of the same form as the input signal. 
%The motivation to consider such a noise  is based on a well-known detection result in linear communication system -- Consider the following detection problem where a receiver aims to estimate the value of an unknown $A$ from the signal $y(t) = A \cdot x(t) +n(t)$. Here, $x(t)$ is a known pulse shape, and $n(t)$ is a white Gaussian noise.  The optimal detector is obtained by passing the signal $y(t)$ through a match filter. In that case, if we decompose the white noise $n(t)$ as the sum $n_{1}(t) + n_{2}(t)$ where $n_{1}$ is a scalar multiple of the input signal $x(t)$ and $n_{2}(t)$ is orthogonal to $n_{1}(t)$, then the perturbation of detected value for $A$ is only affected by $n_{1}(t)$, but not by $n_{2}(t)$. In other words, in the detection problem above, the noise $n_{1}(t)$ contributes solely on the perturbation. Therefore, in the following, we want to examine if the noise $n_{1}(t)$ still plays a similar and significant role in the perturbation of eigenvalues or not.
We assume the input to the segment is $q(t) $ and $n(t)$ is the white Gaussian noise added in the segment. Define $n_{1}(t)$ and $n_{2}(t)$ such that 1) $n(t) = n_{1}(t) + n_{2}(t)$, 2)  $n_{1}(t)$ and $n_{2}(t)$ are orthogonal to each other, and 3) $n_{1}$ is a scalar multiple of $q(t)$. 
Notice that the power of $n_{1}(t)$ is significantly smaller\footnote{
Strictly speaking, for white noise, $n_{2}(t)$ has infinite power while $n_{1}(t)$ has finite power. }  than $n_{2}(t)$. We will call $n_{2}(t)$ the residual noise and $n_{1}(t)$ the \emph{scaling noise} -- adding $n_{1}(t)$ to $q(t)$ is equivalent to multiplying $q(t)$ by a scaling factor. In the following numerical example, we will evaluate the impact of scaling noise and residual noise on discrete eigenvalues. 

\begin{figure}
\centering
\includegraphics[width=9cm]{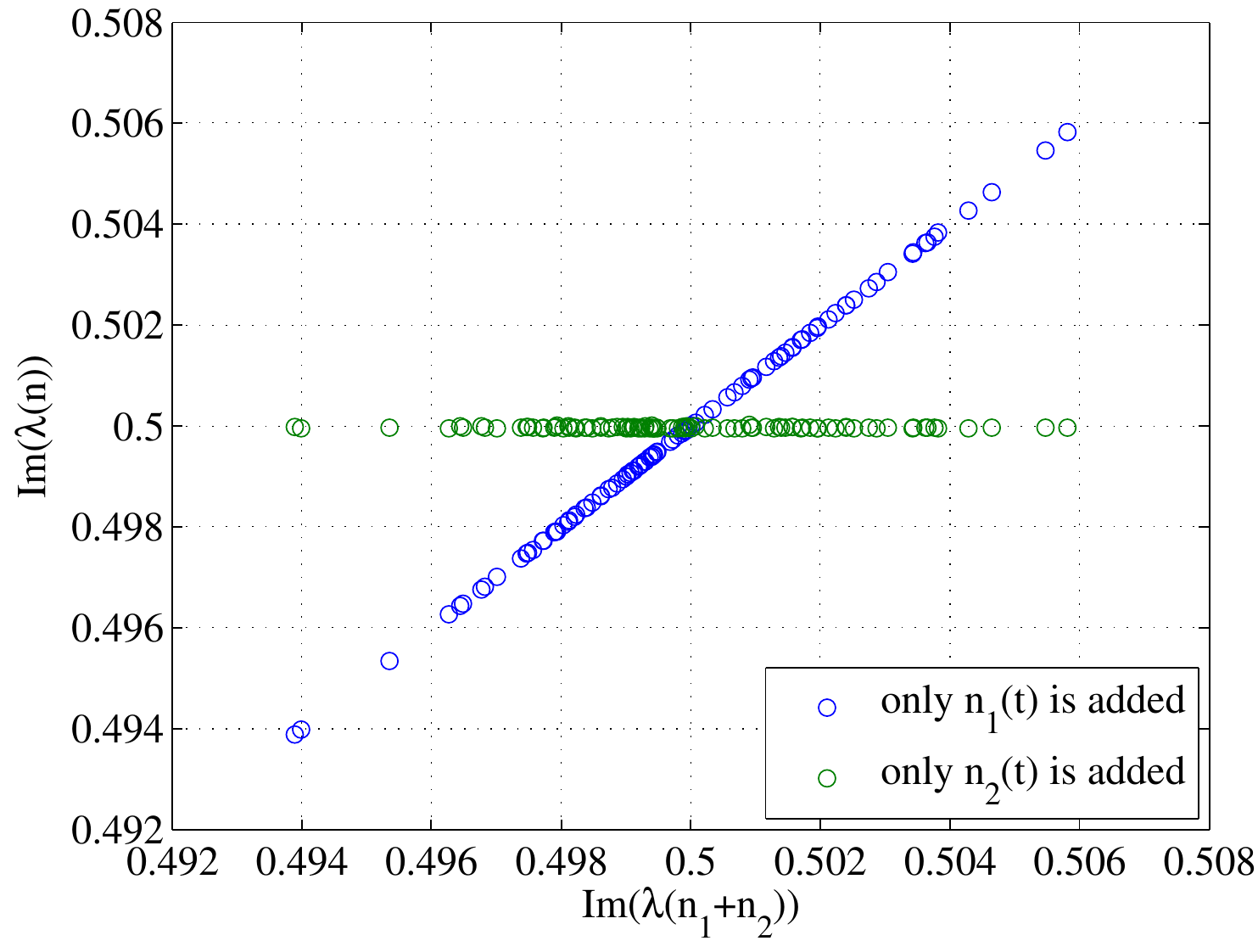}
\caption{Eigenvalue perturbation errors when $n_{1}(t)$ and $n_{2}(t)$ are added separately.}
\label{fignoisedominate}
\end{figure}

\begin{figure}
\centering
\includegraphics[width=9cm]{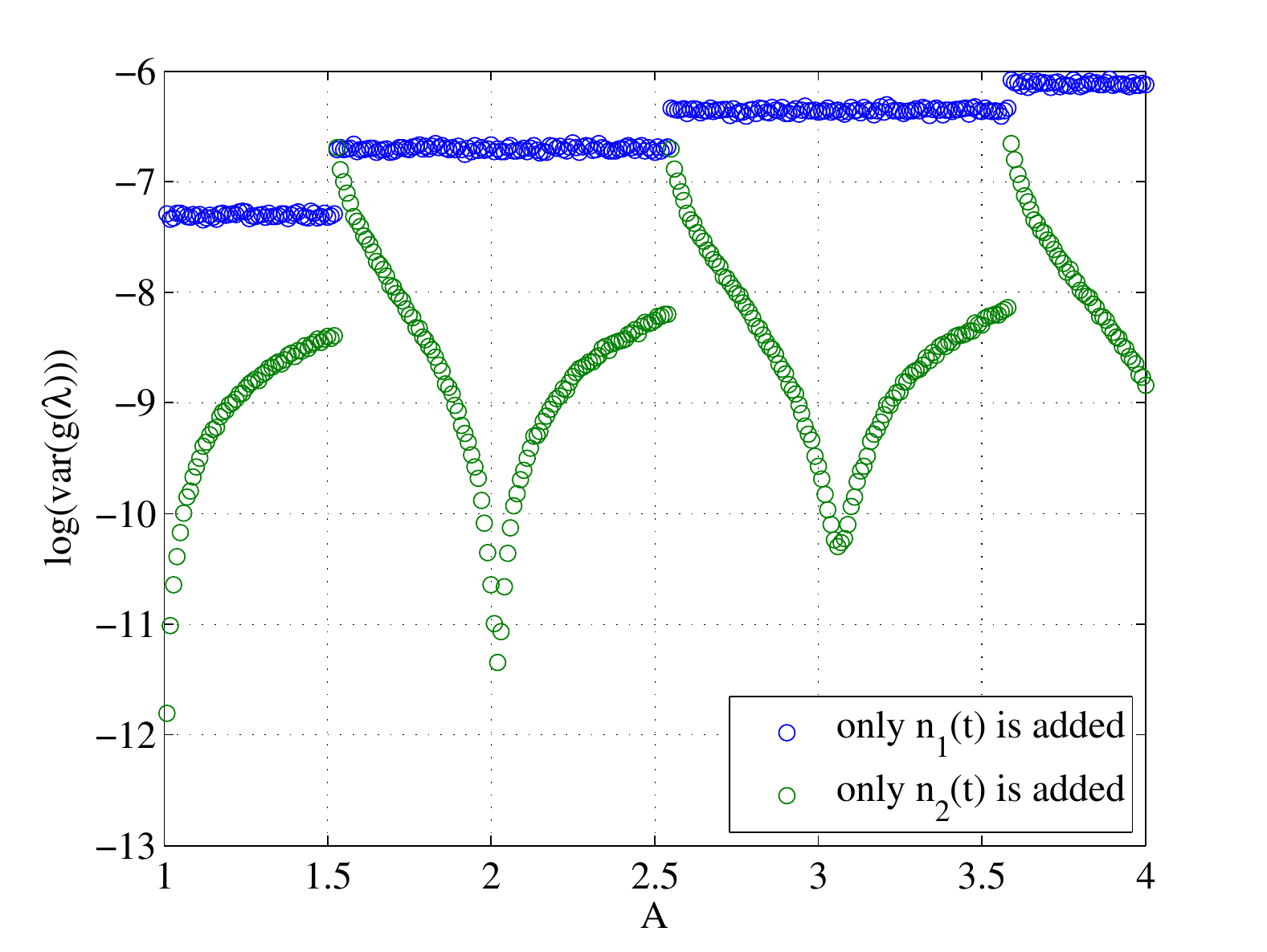}
\caption{Variances of eigenvalue perturbation errors when $n_{1}(t)$ and $n_{2}(t)$ are added separately}
\label{figNoiseCQ}
\end{figure}

First, we let the signal $q(t)$ be a fundamental soliton. And we will compare the imaginary part of the discrete eigenvalues of  $q(t) + n_{1}(t)$, $q(t) + n_{2}(t)$ and $q(t) + n_{1}(t)+n_{2}(t)$.
Results are shown in Figure \ref{fignoisedominate} and  \ref{figNoiseCQ}.
In Figure \ref{fignoisedominate}, we consider $q(t)$ as the fundamental soliton first. The $x$-axis denotes the imaginary part of the discrete eigenvalue of   $q(t) + n_{1}(t)+n_{2}(t)$
and the $y$-axis corresponds to that of $q(t) + n_{1}(t)$ and $q(t) + n_{2}(t)$.  
The figure clearly shows that when only $n_{2}(t)$ is added to the signal, the eigenvalue is largely unchanged (see the green circles data points). On the other hand, when only $n_{1}(t)$ is added, the eigenvalue is essentially the same as the one obtained by adding both noises together. This example illustrates that the scaling noise is dominating the perturbation of discrete eigenvalues. 
 
Next, we consider the case when $q(t) = A \text{sech}(t)$. In this case, $q(t)$ has at least two eigenvalues $A-0.5$ and $A-1.5$ for   $A > 1.5$.
%has more than one eigenvalue, the situation becomes more complicated, as one needs to consider perturbations of multiple eigenvalues. We notice that the perturbation of individual eigenvalues is much higher than the  sum of the eigenvalues. Therefore, 
In our example, we focus only on the sum of the imaginary parts of the discrete eigenvalues  (which can be interpreted as the amount of energy in the solitonic component of the signal).
%Let 
%\[
%g(\lambdav)  = \text{imag}\left(\sum_{i} \lambda_{i} \right)
%\]
%where $\lambdav = (\lambda_{i} , i= 1, \ldots, |\lambdav|)$.
%We observe something very similar that the noise $n_{1}(t)$ seems to dominate the perturbation of the sum of eigenvalues, compared to the other noise $n_{2}(t)$. 
%As an illustration, we consider the case when 
%For each $A$, we consider the variances of the sum of the eigenvalues, when only $n_{1}(t)$ and $n_{2}(t)$ are added. 
Our simulation shows that the scaling noise $n_{1}$ has a more significant impact on the perturbation (measured by variances) of the sum of eigenvalues. Specifically,  we notice that

\begin{enumerate}

\item 
Eigenvalue perturbations caused by addition of $n_{1}$ are often much bigger than that by addition of  $n_{2}$;
 
\item Impact caused by $n_{2}$ on eigenvalue perturbation is at the smallest  when $A$ is close to an integer (i.e., when $q(t)$ is a multi-soliton) and is at the largest when $A$ is slightly greater than $A-0.5$  is slightly bigger than an integer (i.e.,  when   $q(t)$ has an eigenvalue  close to zero) 

\item Impact caused by $n_{1}(t)$ is constant over regimes when $q(t)$ has the same number of discrete eigenvalues.
\end{enumerate}

%\subsubsection*{Implications}
%In the above examples, we observe that the noise component $n_{1}$ (which matches the signal $q(t)$) seems to play a significant role in characterising the perturbations of the discrete eigenvalues. 

Motivated by our observation, we propose the following simplification: 
According to our previous model, eigenvalues perturbation can be approximated by the sum of a collection of local perturbations 
$
g(\hat \lambdav_{m}) - g( \lambdav_{0})
$
where $\hat \lambdav_{m}$ are discrete eigenvalues  of  
$
\hat q_{m}(t) \triangleq \bar q_{m}(t) + n_{m}(t).
$

Let $n_{m}^{(1)}(t)$ be the scaling noise component of  $n_{m}(t)$, 
\[
{\hat{\hat q}}_{m}(t) \triangleq \bar q_{m}(t) + n^{(1)}_{m}(t)
\]
and   ${\hat{\hat {\lambdav}}}_{m}$ be its corresponding set of discrete eigenvalues. Then we can approximate $\hat \lambdav_{m}$ with ${\hat{\hat {\lambdav}}}_{m}$.  

{\bf Remark: }
The merit of the simplification is that it is analytically simpler as $n^{(1)}_{m}(t)$ is a one dimensional noise with finite power.

\begin{figure}
\centering
\includegraphics[width=8.5cm]{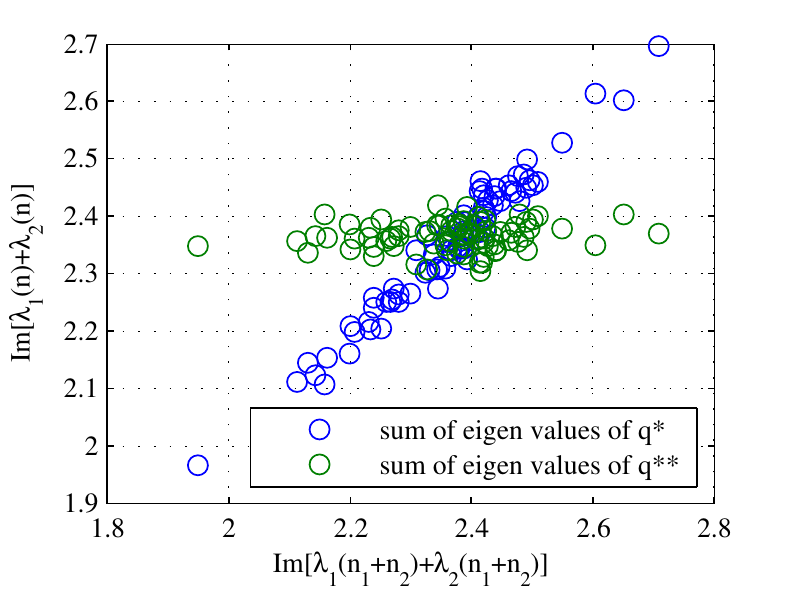}
\caption{Noise decomposition for 2-soliton with discrete eigenvalues $0.9 \imag$ and $1.5 \imag$. }
\label{noisedecompose2a}
\end{figure}

Experimental observations confirm the above claim. 
% In Section \ref{sec:noisedecomposition}, we propose to decompose noises into two components and claim that the ``scaling noise component'' contributes significantly to the perturbation of discrete eigenvalues. Our observation is supported by simulations. In the following, we will conduct experiments to support our claim. 
Consider a 2-soliton input signal with discrete eigenvalues $0.9 \imag$ and $1.5 \imag$. 
Figure \ref{noisedecompose2a} plots the sum of the imaginary parts of the   two eigenvalues of $q_{M}(t), q^{*}$ and $q^{**}$ as defined by; 
1) $q_{M}(t) = q_{0}(t) + n_{1}(t) +n_{2}(t)$, 2) $q^{*}(t) = q_{0}(t) + n_{1}(t)$ and 2) $q^{**}(t) =  q_{0}(t) + n_{2}(t)$. In Figure \ref{noisedecompose2a}, the $x$-axis denotes the (imaginary part) of the sum of the  eigenvalues of $q_{M}(t)$, while the $y$-axis denotes that of $ q^{*}$ and $q^{**}$. It is observed that the perturbation of eigenvalues of $q_{M}(t)$ is largely contributed by the noise $n_{1}(t)$.

%In addition, ignoring the residual noise, the channel is less hostile and hence 
%
%
%Furthermore, the channel corresponding to this further eigenvalue perturbation approximation model is less hostile, as some of the noises are ignored. In this case, upper bound on capacity  obtained using this model  will remain valid under the original model. 
% 
% 

%\subsection{Experimental examples}

%So far, we have proposed a model  for  discrete eigenvalues perturbation by ``decomposing'' the perturbation as a sum of smaller perturbations, each caused by noise added in a short fibre segment.  We then further demonstrate  that a significant portion of the perturbation is in fact caused the scaling noise.  

%%%%%%%%%%%%%%%%%%%%%%%%%%%
\section{{\bf Methods}}
% \section{{\bf Results:} Modeling eigenvalues perturbation}
 \subsection{\bigskip Simulation}
\label{Sim}
 Numerical NFT was implemented based on the work in
 \cite{frank1}. Forward difference method has been
 selected. This method recursively calculates eigenvector $\nu$ from initial
 condition  (\ref{ch3.2.9}). The time interval of the input
 signal $[T_{1},T_{2}]$ is divided into $N$ steps with each step size $(T_{2}-T_{1})/N$. The
 initial condition is:
 \begin{equation}
 \nu(T_{1},\lambda)=\left(
 \begin{array}
 [c]{c}%
 1\\
 0
 \end{array}
 \right)  e^{- \imag\lambda T_1}.
 \end{equation}
 Once the final value i.e. eigenvector $\nu(T_2,\lambda)$ is found by recursive
 processing it is inserted in following equations to find $a(\lambda)$ and
 $b(\lambda)$:
 \begin{equation}
 a(\lambda)=\lim\limits_{t\rightarrow\infty}\nu_{1}(t,\lambda)e^{ \imag  \lambda t}%
 \end{equation}%
 \begin{equation}
 b(\lambda)=\lim\limits_{t\rightarrow\infty}\nu_{2}(t,\lambda)e^{- \imag\lambda t}%
 \end{equation}
 where $\nu(T_2,\lambda)\triangleq(\nu_{1}(T_2,\lambda),\nu_{2}(T_2,\lambda))^{T}$.

 For the discrete spectrum, it is required to find the values of $\lambda$ for
 which $a(\lambda)$ becomes zero. This procedure was performed by creating user
 defined function in MATLAB that takes initial guess value of $\lambda$ and uses
 standard functions in MATLAB to find all $\lambda$s corresponding to $a(\lambda)=0$.
 MATLAB calls that user defined function each time until it becomes zero.

 The NFT code developed using the forward difference table was put to test to
 find eigenvalues of multiple functions e.g. fundamental soliton, non
 fundamental solitons, arbitrary signals and then was checked for error
 percentage. The error percentage was negligibly small $10^{-3}$ when number of
 steps $N > 400$. The implemented code was checked against a few typical
 pulses with known eigenvalues.

 Numerical Nonlinear Pulse Propagation (NPP) based on split-step Fourier method has been widely used for
 simulations of nonlinear optical processes in wave guides. It is proven to have
 high accuracy in predicting pulse evolution during propagation. We used an in-house
 NPP software, written in C/C++,
 and used CUDA to parallelize the calculations by utilizing the power of
 graphic processing unit (GPU). This has reduced the software run-time significantly, allowing
 simulating a large number of instances required for statistical evaluation. 
 In our NPP, the propagation of a pulse along a fiber is divided into
 length segments within which, the nonlinear and linear processes can be separated as an
 approximation. In each step, a band-limited white Gaussian noise is added in
 the frequency domain across the whole spectrum. To limit the noise bandwidth
 to a certain value we have used a band pass filter in the frequency domain. In
 order to accurately describe the statistics of the model, the same propagation
 simulation has been repeated for 5000 times, and eigenvalues were calculated
 using our numerical NFT.

%  Using CPU based pulse propagation code to simulate 5000 pulse propagation for
%  each set of parameters is very time consuming. In order to increase the
%  simulation efficiency, we used a CUDA supported nonlinear pulse propagation
%  solver (NPPS) to speed up the simulation \cite{??}. The NPPS was written in C/C++,
%  and used CUDA to parallelize the calculations and by utilizing the power of
%  graphic processing unit (GPU). This allowed for simulation of the propagation
%  in the order of hours instead of days.

%  In the theory shown in the previous section, the noise is modelled as white
%  Gaussian noise. However, in numerical simulation, it is impossible to simulate
%  infinite broadband white noise. However, we used the full spectral range that
%  available in the Fourier window in the NPP. To confirm, the limited bandwidth
%  do not have significant impacted on the results of this study, we first study
%  the effect of noise bandwidth on NFT.

%  \bigskip

%  NFT has been studied
%  \cite{1993HN,DBLP:journals/corr/abs-1202-3653,DBLP:journals/corr/abs-1204-0830,DBLP:journals/corr/abs-1302-2875}
%  and fast NFT \bigskip algorithms have beed developed
% %  \cite{201305WP,2015WPearly,\boldsymbol{\Phi}}.

%  \bigskip

 \subsection{\bigskip Experiment}

\subsubsection{Transmission setup} The drive signals were converted to the analog domain by
 high-speed AWG (Keysight M8196A) operating at up to 92 GSa/sec. The lasers
 (both carrier and LO) used in the experiments external cavity lasers (ECL)
 emitting near 1,550 nm with a linewidth of \symbol{126}100 kHz. The modulators
 used were Mach-Zehnder I/Q modulators based on LiNbO3 waveguides. A 50km
 Non-zero dispersion-shifted fiber (NZ-DSF) with a nonlinear Kerr coefficient
 $\sim 1.2 W^{-1} km^{-1}$, a dispersion coefficient of $\sim 4 ps\cdot nm^{-1}km^{-1}$,
 and 9.5 dB insertion losses was chosen as the transmission medium in the
 fiber loop. In this case, the distance of 150km in a normalized NLS was around
 0.1, which was obtained through the variable conversion shown in [3].
 Before launched into the fiber loop, the launch powers were carefully controlled by the attenuator after a fixed gain EDFA (with noise figure 5dB) to the optimum value. One extra EDFA was used to compensate for the remaining loss in the loop, and a flat-top optical filter with a 3dB bandwidth of 1nm is followed to suppress the out-of-band amplified spontaneous emission(ASE) noise. At the receiver, a polarization controller is used to align the optical signal in the x-polarization. Then the signal was detected by a dual polarization optical coherent receiver consists of a 90%
 %TCIMACRO{\U{b0} }%
 %BeginExpansion
 ${{}^\circ}$
 %EndExpansion
 hybrids and  4 balance  pin-photodetector with 3dB bandwidth of \symbol{126}38GHz . The output 4 E-fields
 waveforms were sampled by a digital storage scope (Agilent 96204Q) with a
 sampling rate of 80GS/s and a bandwidth of 33GHz and stored to process offline.

 \subsubsection{Transmitter and receiver DSP} At the transmitter, the various 2-soliton pulses
 were recursively computed using the Darboux transformation method [1]. The
 initialization coefficients $A_{i}$ and $B_{i}$ for Darboux transformation method which
 (the discrete-spectral amplitudes and the shape of the signal) were specially
 choose to get smaller physical bandwidth to improve the performance at
 transmitter [2]. The receiver DSP firstly used a training symbol to perform
 timing synchronization. Then A pilot tone from y-polarization was used to estimate and compensate the laser phase noise and frequency offsets. After normalized by a scaling factor according to
 the lossless path averaged model, the synchronized pulse train was processed
 per pulse to Search the corresponding roots. Ablowitz-Ladik algorithm
 was used to calculate the Nonlinear Fourier Coefficients, followed by
 a Newton-Raphson method for root searching [5].

\section{Further Works and Conclusions}
This paper focuses on perturbations/noises of eigenvalues when the optical signal is transmitted along a fibre. We have numerically and experimentally demonstrated that the noises are correlated. By exploiting the correlation, one can design a better signal constellation leading to a higher system throughput. 
In order to take advantage of the correlation, it becomes important to derive a model of the eigenvalue noises. In the second part of the paper, we have proposed an analytical framework to characterise the noises. The idea is to decouple the eigenvalue perturbation as an accumulation of many smaller perturbations, each of which is caused by the addition of noises in a short fibre segment. As a result, one can derive an eigenvalue perturbation model by characterising each smaller perturbations. Strictly speaking, all of these small perturbations are non-identically distributed and are also correlated with each other. However, we observe that the correlation is indeed quite weak that one can essentially assume them to be independent. Following the independence relation, the perturbation in eigenvalue caused by propagation can be modeled as the accumulation of a set of independently added noise.

So far, our focus is on the eigenvalue perturbation caused by noises during signal propagation.  However, our modeling framework can also be extended to included noised introduced at the transmitter and the receiver. Specifically, we will model that an additive transmitter/receiver noise will be added respectively before and after the transmission.  When the transmitter and receiver noises are introduced, the signal's eigenvalues will also be perturbed. Following our paradigm, we can model the perturbations introduced at the transmitter and receiver as independent noises. 
Furthermore, we can also extend our work to model the perturbation of spectral amplitudes. Some preliminary correlation studies were done and can be found in \cite{gui_alternative_2017}. In this current paper, our focus is on the perturbation of the eigenvalues. However, the same principle will also apply to spectral amplitudes as well where the overall noises in spectral amplitudes will be modeled as the accumulation of many independently distributed noises caused by the injection of noises in a short fibre segment.

\bibliographystyle{IEEEtran}
%\bibliography{bibtex/bib/References}


%%%%%%%%%%%%%%%%%%%%%%%%%%%%%%%%%%%%

\begin{appendices}
\section{Supplementary materials}
\subsection{Effect of different noise on eigenvalue distribution}
\label{NoiseType}
In order to investigate collectively the effects of noise at different stages of pulse generation, we consider the most general form of noise (which can be various kinds of amplitude, phase and background noise) as follows: 
\begin{equation}
	\left[A_{0}+A_{}(t)\right]q(t)e^{2\pi*i*\left[B_{0}+B_{}(t)\right]}+C_{0}+C_{}(t),
\end{equation}
where $A_{0}, A_{}(t), C_{0}, C_{}(t)$  are complex, and  $B_{0}$ and $B_{}(t)$  are real. We study the effect of different types of  noise by allowing noise in each of $A_{0}, A_{}(t), C_{0}, C_{}(t), B_{0}$ and $B_{}(t)$ separately. The following figures (\ref{figNoiseType}) show the NFT eigenvalue distributions for the 2-eigenvalue pulses (the same eigenvalue sets as the system described in Fig. \ref{figExpResult} in the main document) when noise is added to each of the coefficients individually. All figures, except Fig. \ref{figNoiseType}(e), represent similar trend; positive correlation and elliptical shape distribution with minor and major axes whose values and orientations\footnote{
For each group of simulation/experiment, a set of 2-dimensional data points (whose coordinates are the discrete eigenvalues of the received signals) will be generated. These data points further defines an empirical covariance matrix. Its orientation is defined as the slope of the principal eigenvector of the covariance matrix.
} 
depend on the values of the set of ($\lambda_1$ and $\lambda_2$).  It is also observed that comparatively constant noises have less impact on the orientation, values of minor and major axes, and shifting of the mean values of each set than those of time-varying noises. Among time-varying  noises, the phase noise, Fig. (\ref{figNoiseType}f) has the most impact in terms of shifting the mean values of each set. Note that constant phase noise, on the contrary, has no effect on the distribution of eigenvalues whatsoever. This agrees with analysis where it is well known that constant phase changes have no impact on discrete eigenvalues (but only on the corresponding spectral amplitudes).  Although the exact contribution each type of the noise is unknown, the difference between the experimental observation and simulation can attributed to these types of noise.

% .  and the correlation  components of  The top row (Fig. \ref{figNoiseType}(a)) shows the effect of amplitude noise. When only $A_{const}$ noise presents, the eigenvalues of the signals are linearly correlated with an average correlation of 0.999. Fig. \ref{figNoiseVarAngle}(a) shows the amplitude (??? what do we mean. change it to correlation) of the noise increases as the $\lambda$ increases. For different eigenvalue combination, the angle of the primary correlation direction is slightly different. A general trend can be found that, with the increase of $\lambda_1$, the angle decreases, and with the increase of $\lambda_2$, the angle increases. Similar phenomena can also be found when $A_{}(t)$ noise is present as shown in Fig. \ref{figNoiseType}(b). However, the average correlation (????) is smaller than the one in Fig. \ref{figNoiseVarAngle}(a). 
%
\begin{figure}[ht]
\centering
\subfloat[$A_{0}$]{\includegraphics[width=4.5cm]{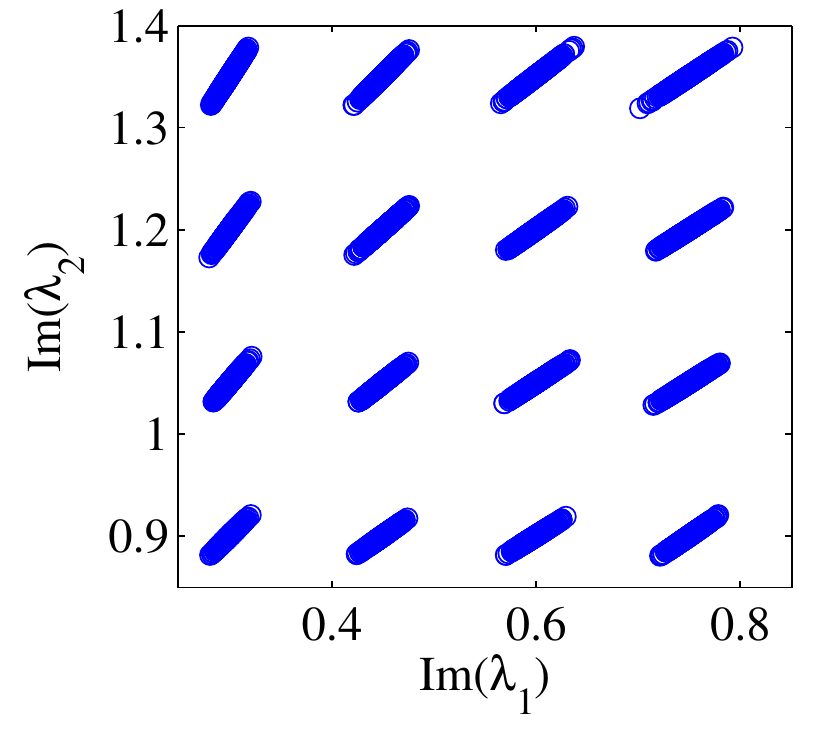}}\subfloat[$A_{}$]{\includegraphics[width=4.5cm]{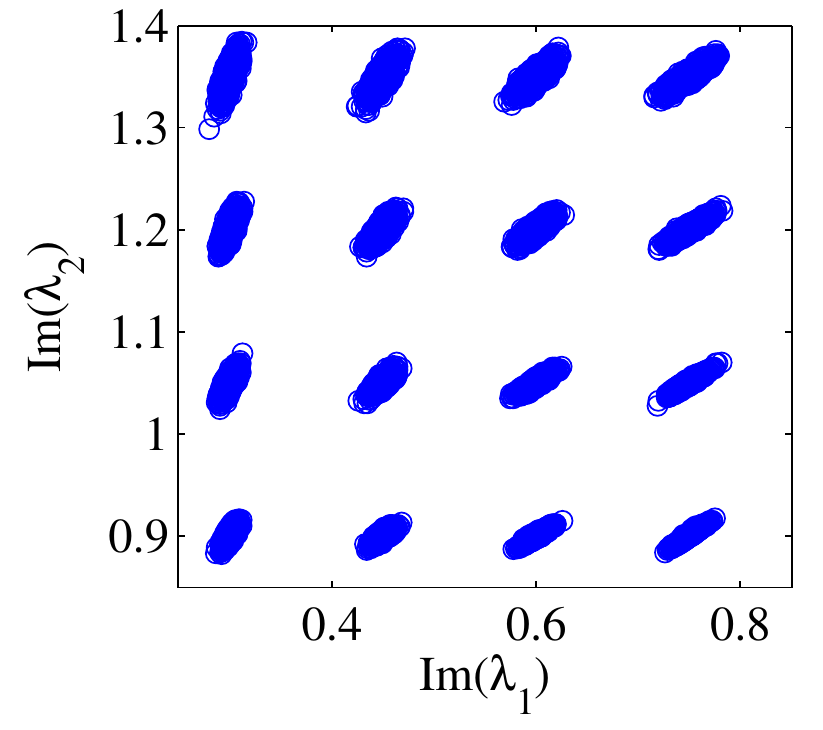}}\\
\subfloat[$B_{0}$]{\includegraphics[width=4.5cm]{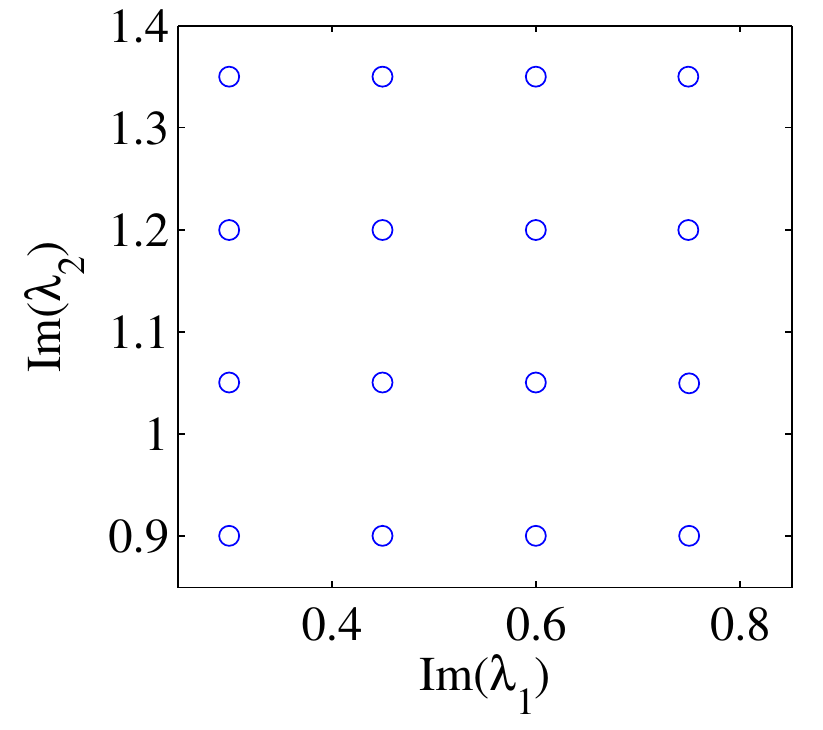}}\subfloat[$B_{}$]{\includegraphics[width=4.5cm]{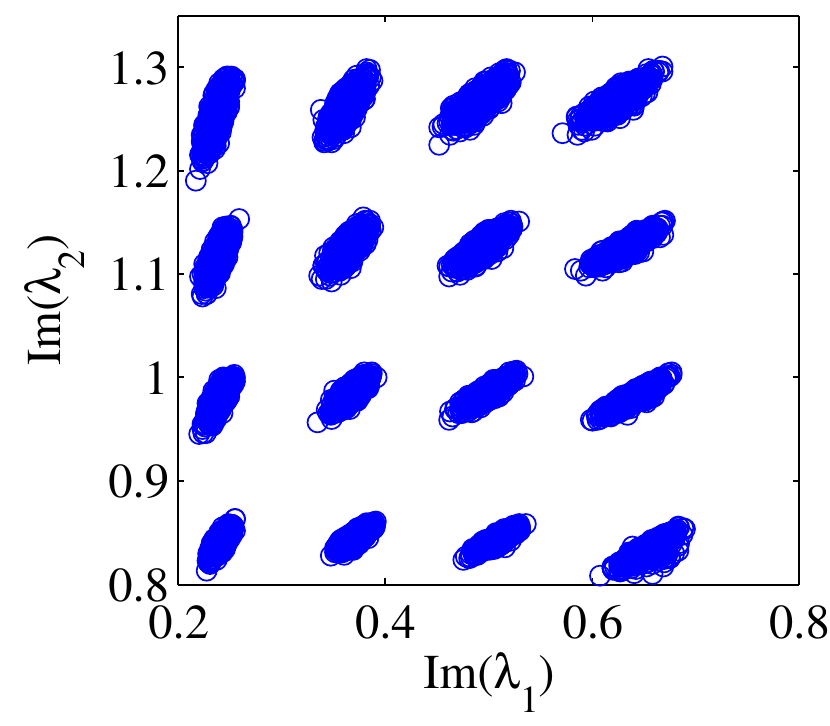}}\\
\subfloat[$C_{0}$]{\includegraphics[width=4.5cm]{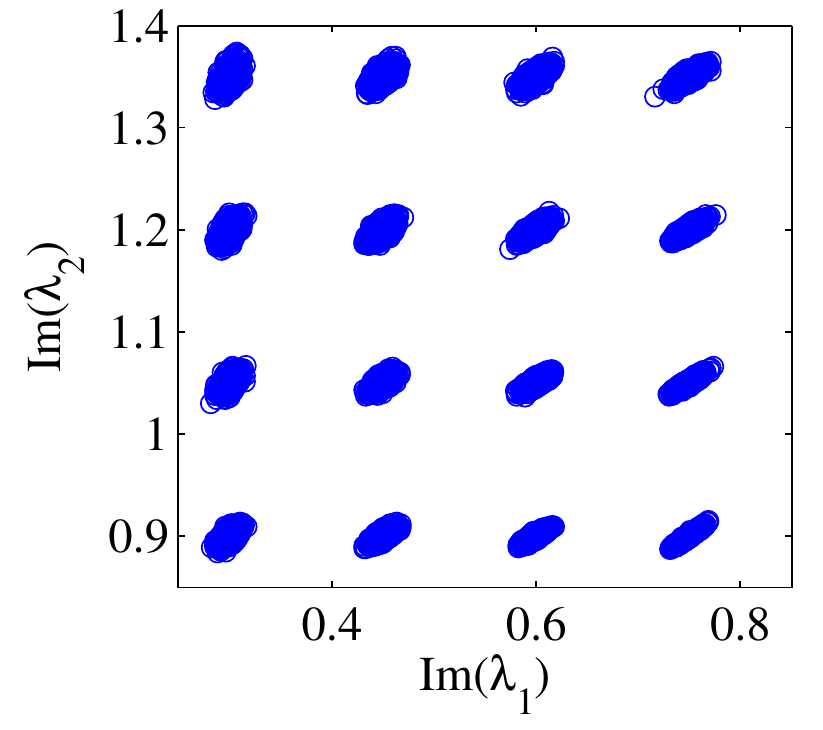}}\subfloat[$C_{}$]{\includegraphics[width=4.5cm]{Fig11C}}
\caption{The effect of different types of noises on NFT eigen values.}
\label{figNoiseType}
\end{figure}

\subsection{Channel noise modeling via linearisation}
In this paper, we proposed a method to model perturbation eigenvalues. As an analogy, our approach is in fact similar to the idea of approximating a nonlinear function with a piece-wise linear function via the approximation that 
%\begin{align}
%f(x_{o} + \delta_{1} + \delta_{2}) - f(x_{o})  & \approx   f'(x_{o})\delta_{1} + f'(x_{o}+\delta_{1})\delta_{2} \label{eq21}\\
%& \approx   f'(x_{o})\delta_{1} + f'(x_{o})\delta_{2} \label{eq22}
%\end{align}
%or that 
\begin{multline}
f(x_{o} + \delta_{1} + \delta_{2}) - f(x_{o})   \\
 \approx f(x_{o} + \delta_{1}) - f(x_{o}) + f(x_{o}  + \delta_{2}) - f(x_{o}) \label{eq23}
\end{multline}
In other words, the error in the function $f$  caused by the noises $\delta_{1}$ and $\delta_{2}$ (either added simultaneously or sequentially) is roughly equal to the sum of errors induced by $\delta_{1}$ and $\delta_{2}$  separately. In the following, we will further elaborate  our results in details.

Our first step is to verify if similar ``linearity'' in \eqref{eq23} also holds or not in our application. Consider the following simple scenario. Let $q_{0}(t)$ be an input signal and 
$\lambdav_{0} = (\lambda_{0,i} , i \in |\lambdav_{0}|)$ 
%$\lambda_{0} = (\lambda^{(i)}_{0} , i \in \Lambda)$ 
be  its corresponding set of discrete eigenvalues and $g(\lambdav_{0})$ be a function of $\lambdav_{0}$ of interest.
Note that $g(\cdot)$ can be a vector-valued function in general. The following are some examples:
\begin{align}
g(\lambdav_{0}) \triangleq \min_{i =1 , \ldots,  |\lambdav_{0}|} \text{imag}(\lambda_{0,i}), 
\end{align}
\begin{align}
g(\lambdav_{0}) \triangleq \max_{i =1 , \ldots,  |\lambdav_{0}|} \text{imag}(\lambda_{0,i}), 
\end{align}
or
\begin{align}\label{eq25}
g(\lambda_{0}) \triangleq \sum_{i =1 , \ldots,  |\lambdav_{0}|} \text{imag}(\lambda_{0,i}).
\end{align}
%
%the smallest, the largest or the sum of the eigenvalues in $\lambda$.
  
Now, consider two noises $n_{1}(t)$ and $n_{2}(t)$. Let 
\begin{align*}
q_{1}(t)  & = q_{0}(t) + n_{1}(t) \\
q_{2}(t)  & = q_{0}(t) + n_{2}(t) \\
q_{3}(t)  & = q_{0}(t) + n_{1}(t) + n_{2}(t).
\end{align*}
The addition of noises will cause the discrete eigenvalue to perturb. 
Specifically, let $\lambdav_{k}= (\lambda_{k,i} , i \in |\lambdav_{k}|)$ be the eigenvalues of $q_{k}$ respectively for $k=1,2,3$. 
Now, the question is to determine whether 
\begin{align}\label{eq21}
g(\lambdav_{3}) - g(\lambdav_{0}) \approx g(\lambdav_{1}) - g(\lambdav_{0}) + g(\lambdav_{2}) - g(\lambdav_{0})
\end{align}
holds or not. 

\begin{figure}[h]
\centering
\includegraphics[width=9cm]{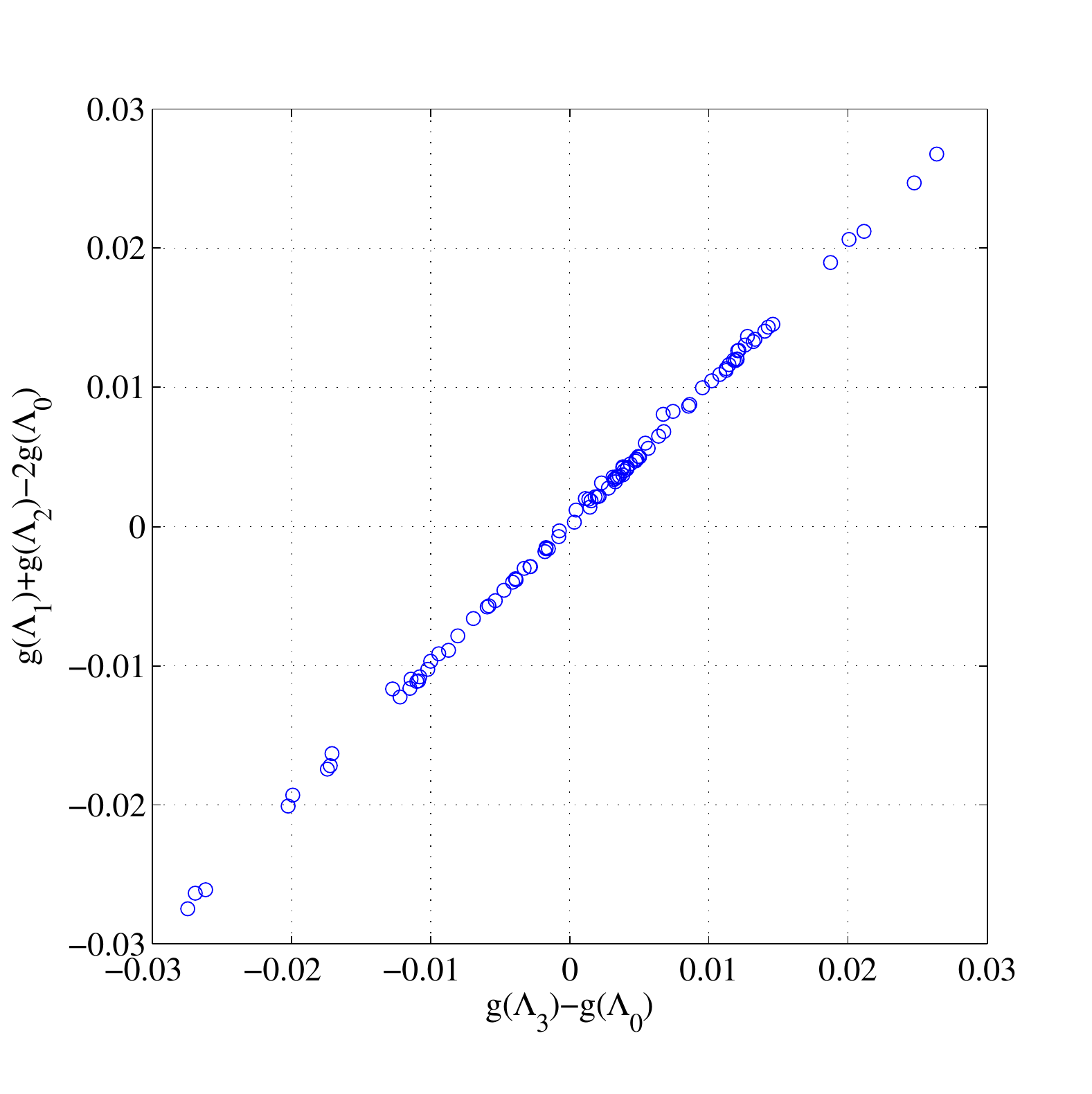}
\caption{Eigenvalue perturbations add up linearly.}
\label{figNoiseLinearity}
\end{figure}

To answer the question, we conduct the following numerical example, in which 
two Gaussian noises ($n_{1}(t)$ and $n_{2}(t)$) are added to a 2-soliton $q_{0}(t)$.
We consider a scalar-valued function $g$ as defined in \eqref{eq25}, which can be interpreted as the energy of  the signal's solitonic components. Results are shown in Figure  \ref{figNoiseLinearity} where the $x$-axis corresponds to $g(\lambdav_{3})-g(\lambdav_{0})$ and the $y$-axis corresponds to $g(\lambdav_{1})   + g(\lambdav_{2}) - 2g(\lambdav_{0})$. Our numerical example clearly indicates that the two quantities are basically the same suggesting that  ``linearity'' in \eqref{eq21} does hold.
The previous example suggests that the errors/perturbations on the eigenvalues are in fact linear -- that the accumulated errors induced by the addition of two noises is approximately the sum of the two errors induced by the noises separately. Next, we will consider the case when the noises are added in a distributed manner along the fibre. 

%However, the above example ignores the fact that noises are not added at a single location but everywhere along the fibre. Therefore, in our next step, we will further take into account the effect of propagation (and fibre nonlinearity) as well. 

\subsection{Transmitter and propagation noise}
\label{TPN}
\begin{figure}[h]
\centering
\includegraphics[width=8cm]{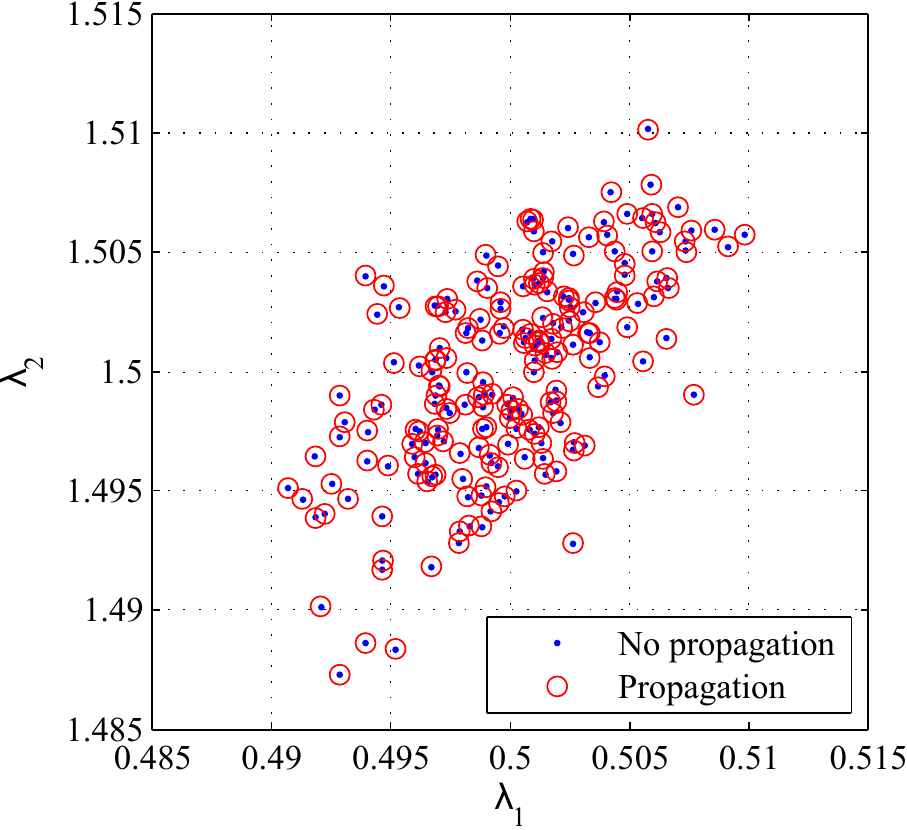}
\caption{Eigenvalue perturbations caused by transmitter noise.}
\label{figTxNoise}
\end{figure}

\begin{figure}[h]
\centering
\includegraphics[width=9cm]{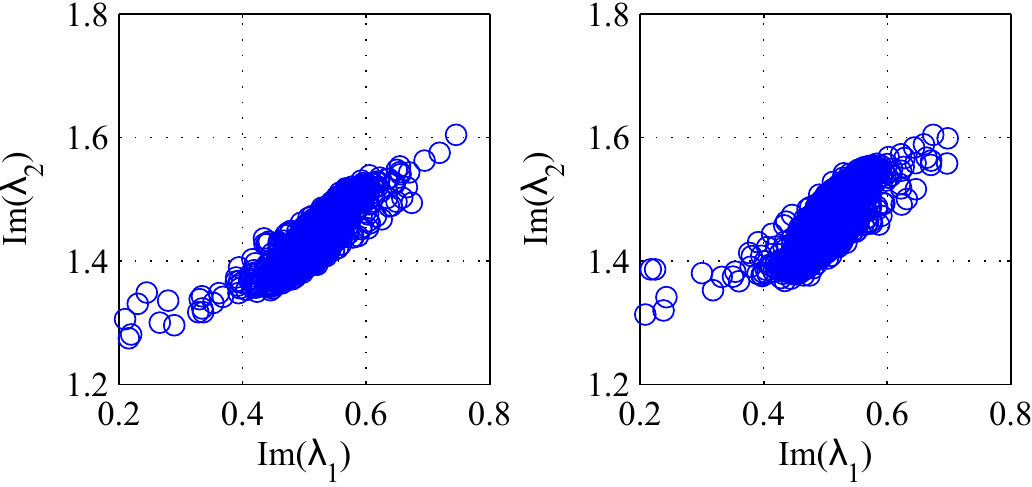}
\caption{Eigenvalue perturbations in experiment. (Left) Back to back pulses. (Right) Pulses after 150 km propagation.}
\label{figPropNoise}
\end{figure}

In section IV B, an experiment was conducted to demonstrate that perturbation of discrete eigenvalues caused by addition of point noise is input dependent (or more precisely, depends on the NFT phase of the pulse in that example). In the experiment, transmitter noise was not considered effecting the outcome and the propagation noise in the fiber is less significant than the receiver noise. To support this assumption, Fig.\ref{figTxNoise} and Fig.\ref{figPropNoise} are produced. Fig.\ref{figTxNoise} is a numerical simulation of a collection of pulses with noise added prior to a noiseless propagation for a distance corresponding to $\pi$ NFT phase (maximum change in the orientation of the covariance matrix). The results indicate the covariance matrix of the pulses after propagation (red circles) is the same as the pulses before propagation (blue dots). In other words, the noise generated in the transmitter does not effect the experiment.

Fig. \ref{figPropNoise} shows the comparison of the covariance matrix of the experimental data sets for 0 km and 150 km of fiber propagation respectively. The variances along the primary axis are 2.48e-3 for 0 km and 2.28e-3 150 km. The covariances are 1.16e-3 for 0 km and 1.01e-3 for 150 km. We consider this amount of change in the variances to be small and thus noise add during the propagation through fiber is small comparing to the noise in the transmitter and the receiver.

In summary, the transmitter noise does not contribute to the changes of the covariance matrix of the received signals. The noise added to the signals during propagation is relatively small. Therefore, the observed rotation of the primary axis of the covariance matrix at different propagation length is mainly due to the noises added at the receiver. Hence, the experiment is equivalent to a point noise added to pulses with different NFT phases.

To see this, first, all the transmitter noise added at the input of the fibre are the same (as they all have the same input pulses). Second, as eigenvalues are invariant under the noiseless propagation, the perturbation of eigenvalues we see at the end of the fibre caused by the transmitter noise will be also be the same as the input. In other words, the noise at the output due to the transmitting stage will be the same for all experiment.

\end{appendices}

\end{document}